\documentclass[11pt,reprint,showkeys, amsmath,amssymb, aps, floatfix]{revtex4-2}
\usepackage{graphicx,subfigure}
\graphicspath{{figures/}}
\usepackage[export]{adjustbox}
\usepackage{dcolumn}
\usepackage[outdir=./epsTopdf/]{epstopdf}
\usepackage{bm}
\usepackage{amsmath}
\usepackage{amssymb}
\usepackage{physics}
\usepackage{xcolor}
\usepackage{slashed,cancel}
\usepackage{multirow}
\usepackage{float}
\usepackage{capt-of}
\usepackage{comment}
\usepackage{ mathrsfs }
\usepackage[colorlinks=true,linkcolor=blue,urlcolor=blue,citecolor=blue]{hyperref}

\begin{document}
	
	\preprint{APS/123-QED}

	\title{Probing top quark anomalous moments in $W$ boson associated single top quark  production at the LHC using  polarization and spin correlation}
	
	\author{Rafiqul Rahaman}
	\email{rafiqulrahaman@hri.res.in}
	\affiliation{Regional Centre for Accelerator-based Particle Physics, Harish-Chandra Research Institute, A CI of Homi Bhabha National Institute,  
		Chhatnag Road, Jhunsi, Prayagraj 211019, India}
	\author{Amir Subba}
	\email{as19rs008@iiserkol.ac.in}
	\affiliation{Indian Institute of Science Education and Research, Kolkata}
	
	\date{\today}
	
	\begin{abstract}
We study the $W$ boson associated single top quark production at the Large Hadron Collider (LHC) to probe anomalous chromo-magnetic and chromo-electric moments of the top quark with the help of polarization and spin correlation observables besides the cross~section in the leptonic final state. We reconstruct the two neutrinos in the final state using the $M_{T2}$ assisted on-shell (MAOS)  reconstruction method to measure the polarization and spin correlation asymmetries of the top quark and the $W$ boson. We estimate the limits on the anomalous moments in a detector-level simulation considering possible backgrounds for a few sets of integrated luminosities and examined the effect of systematic uncertainties. 
\end{abstract}
	
	
	\maketitle
	
    \section{Introduction}\label{sec:intro}
\begin{figure}[!h]
    \centering
    \includegraphics[scale=0.65]{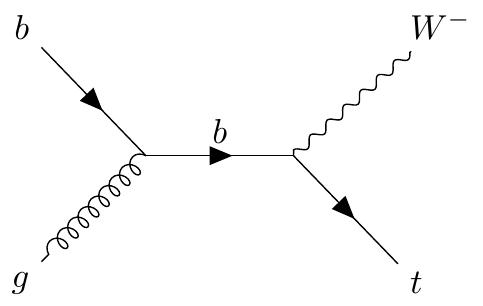}
    \includegraphics[scale=0.85]{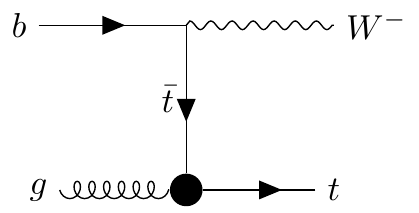}
    \caption{Leading order Feynman diagrams representing $tW^-$ production. The blob in the $t$-channel diagram in {\em right-panel}  represents the presence of anomalous contribution to the $gt\bar{t}$ vertex.}
    \label{fig:feyndiag}
\end{figure}

The top quark being the heaviest known elementary particle in the Standard Model (SM), its coupling strength to the Higgs boson is stronger than the other  SM particles, and it thus offers a natural testing ground for any beyond the SM physics (BSM) that affect electroweak (EW) symmetry breaking. The top quark gets produced copiously in the pair production process through $q\bar{q}$ and  $gg$ initial state. However, the single top quark production in association with a quark (including $b$-quark) or a $W$ boson is significant enough to be observed at the high energy available at the Large Hadron Collider (LHC). The single top quark production in association with a  $W$ boson  ($tW\equiv tW^-+\bar{t}W^+$) process has been observed at the LHC~\cite{RodriguezBouza:2019kbi,CMS:2018amb,ATLAS:2015igu,CMS:2014fut,ATLAS:2016ofl,CMS:2012pxd,ATLAS:2012bqt} and the measured production cross~section matches well with the existing state of the art theoretical estimate with higher order corrections in quantum chromodynamics (QCD)~\cite{Zhu:2002uj,Campbell:2005bb,Re:2010bp,Kidonakis:2010ux,Kidonakis:2021vob} and electroweak~\cite{Frixione:2008yi}. The leading order (LO) Feynman diagrams for the  production  $tW$ process are represented in Fig.~\ref{fig:feyndiag}. Besides top quark pair ($t\bar{t}$) production, the $tW$ production process provides a complementary channel to test for BSM physics connecting to top quark.  In this article, we intend to probe any deviation to the $gt\bar{t}$ couplings in $tW$ production as complementary to the $t\bar{t}$ production process~\cite{CMS:2019kzp,Haberl:1995ek,Atwood:1992vj,Cheung:1995nt,CMS:2016piu,Bernreuther:2013aga,Haberl:1995ek,Hioki:2009hm,Hioki:2013hva,Barducci:2017ddn,Cheung:1995nt,HIOKI:2011xx,Kamenik:2011dk,Aguilar-Saavedra:2014iga,BuarqueFranzosi:2015jrv,Hesari:2012au,Choudhury:2009wd,Gupta:2009eq,Gupta:2009wu,Fabbrichesi:2013bca,Biswal:2012dr}. Deviation to the $gt\bar{t}$ coupling produces anomalous chromo-magnetic dipole moment (CMDM) and chromo-electric dipole moment (CEDM) of the top quark. These anomalous moments are zero at the tree level in the SM, and they receive nonzero small contributions coming higher order effects. The CMDM arises at the one-loop level from both the quantum chromodynamics (QCD) and (EW) sectors~\cite{Aranda:2020tox,Choudhury:2014lna,Bermudez:2017bpx,Martinez:2007qf}. On the other hand, the CEDM  appears only at three-loop levels, arising from the complex phase in the Cabibbo-Kobayashi-Maskawa~(CKM) matrix~\cite{Czarnecki:1997bu,Khriplovich:1985jr}.

The $gt\bar{t}$ interaction Lagrangian, including possible deviation, is generally parameterized in a model-independent way as~\cite{Haberl:1995ek,CMS:2019kzp},
\begin{align}\label{eq:Laggtt}
\mathscr{L}_{gtt} = -g_s\bar{t}\gamma^\mu G^a_\mu \frac{\lambda^a}{2}t 
-\frac{g_s}{2m_t}\bar{t}\sigma^{\mu\nu}\left(\hat{\mu}_t+i\hat{d}_t\gamma_5\right)G^a_{\mu\nu}\frac{\lambda^a}{2}t.
\end{align}
Here $\hat{\mu}_t$ and $\hat{d}_t$ are the top quark CMDM and CEDM, respectively; $\lambda$ are the Gell-Mann matrices; $G_{\mu\nu}$ is the gluon field strength tensor; $g_s$ is the strong coupling constant;  $m_t$ is the top quark mass. The couplings $\hat{\mu}_t$ and $\hat{d}_t$ are $CP$-even and odd, respectively.  

The deviation to $gt\bar{t}$ can also be parameterized by higher dimensional effective operators constructed from the SM fields. At the lowest order, the anomalous $gt\bar{t}$ couplings receive contribution from the dimension-$6$ operator~\cite{Buchmuller:1985jz,Aguilar-Saavedra:2018ggp,Grzadkowski:2010es,Aguilar-Saavedra:2018ksv,Aguilar-Saavedra:2008nuh}
\begin{equation}\label{eq:operator}
\mathscr{O}_{uG\Phi} =\left(\bar{t}\lambda^a\sigma^{\mu\nu}t\right)\tilde{\Phi}G^a_{\mu\nu},
\end{equation}  
where $\Phi$ is the Higgs doublet and $\tilde{\Phi} = i\tau_2\Phi^\star$. The anomalous couplings $\hat{\mu}_t$ and $\hat{d}_t$ in Eq.~(\ref{eq:Laggtt}) are related to the Wilson coefficients (WC) of the operator $\mathscr{O}_{uG\Phi}$ as~\cite{Aguilar-Saavedra:2008nuh},
\begin{align}\label{eq:lag-vs-eft}
\hat{d}_t &= \frac{1}{\sqrt{2}m_t}\text{Im}(C_{uG\Phi})\frac{vm_t}{\Lambda^2}, \notag\\
\hat{\mu}_t &= \frac{1}{\sqrt{2}m_t}\text{Re}(C_{uG\Phi})\frac{vm_t}{\Lambda^2},
\end{align}
where $C_{uG\Phi}$ is the WC, $\Lambda$ is some cut-off energy scale, and $v=246$ GeV is the vacuum expectation value~(VEV) of the Higgs field.
\\

The anomalous $gt\bar{t}$ couplings have been studied earlier in $t\bar{t}$ production process~\cite{Bernreuther:2013aga,Haberl:1995ek,Hioki:2009hm,Hioki:2013hva,Barducci:2017ddn,Cheung:1995nt,HIOKI:2011xx,Kamenik:2011dk,Aguilar-Saavedra:2014iga,BuarqueFranzosi:2015jrv,Choudhury:2009wd,Hesari:2012au,Gupta:2009eq,Gupta:2009wu,Fabbrichesi:2013bca,Biswal:2012dr}, in $tW$ process~\cite{Rizzo:1995uv,YaserAyazi:2013vxz,Fabbrichesi:2014wva}, and also in $t\Bar{t}h$ process~\cite{Cirigliano:2016njn,Cirigliano:2016nyn}.  On the experimental side, these couplings have been probed in various experiments~\cite{CMS:2016piu,CMS:2019nrx,CMS:2019kzp}. A recent study by CMS~\cite{CMS:2019kzp} in $t\bar{t}$ production process provide tightest constraint on  CMDM to be,
\begin{equation}
-0.044 < \hat{\mu}_t < +0.005,
\end{equation}
and for CEDM to be,
\begin{equation}
|\hat{d}_t| < 0.03,
\end{equation}
at $95\%$ confidence level (C.L.). In addition to anomalous $gt\Bar{t}$ coupling, the $tW$ production process have also been studied to probe various other new physics phenomenon~\cite{Tait:2000sh,Cao:2007ea,Barger:2009ky,Rindani:2011pk,Cao:2015doa,Fabbrichesi:2014wva,Aguilar-Saavedra:2008quj,CMS:2019zct}.

The top quark and the $W$ boson being massive, their polarizations and spin correlations are transferred to their decay products' angular distributions~\cite{Boudjema:2009fz,Rahaman:2021fcz}. The polarizations and spin correlations are sensitive to any BSM physics affecting the production mechanism.  
In this article, we intend to probe the anomalous $gt\bar{t}$ couplings in the leptonic final state of  $tW$ production process using the polarization and the spin correlation observables of the top quark and $W$ boson. We note that polarization and spin correlation has been used earlier to probe $gt\bar{t}$ couplings in the $t\bar{t}$ production process~\cite{Rindani:2015vya,Biswal:2012dr,Cheung:1995nt,CMS:2016piu,Bernreuther:2013aga}. The polarization of the top quark has also been studied in single top quark production processes to understand SM background~\cite{Mahlon:1999gz,Boos:2002xw,Espriu:2002wx}. Possible deviation to $tbW$ vertex has been studied in the $tW$ process using polarization of top quark and $W$ boson~\cite{Deliot:2017byp,Rindani:2015vya,Rindani:2011pk,PrasathV:2014omf,CMS:2020ezf}. For simplicity, we neglect the contribution of anomalous $tbW$ couplings in our study. 

For the measurement of polarization and spin correlation, one has to study the angular distributions of the decay products at the rest frame of the top quark and the $W$ boson. This requires one to reconstruct the top quark and $W$ boson four-momenta, which requires the reconstruction of two missing neutrinos in the final state. We use the collider variable $M_{T2}$~\cite{Barr:2003rg,Lester:1999tx} assisted on-shell~(MAOS)~\cite{Cho:2008tj} reconstruction technique to reconstruct the four-momenta of the missing neutrinos. 

The rest of the article is arranged as follows. 
In Sec.\ref{sec:pol-corr}, we discuss the formalism for the polarization and spin correlation in the $tW^-$ production process.  In Sec.~\ref{sec::mt2}, we discuss reconstructing the momenta of two missing neutrinos using the MAOS algorithm, and in Sec.~\ref{sec::probe}, we probe the anomalous couplings and study the effect of luminosity and systematic uncertainties. We conclude in Sec.~\ref{sec::con}.
\section{Polarizations and spin correlations}
\label{sec:pol-corr}
The polarizations of the top quark and $W$ boson, including their spin correlations, can be parameterized in a spin-correlated density matrix given by~\cite{Rahaman:2021fcz}
\begin{eqnarray}\label{eqn:dm-pocorr}
P^{tW}(\lambda_t,\lambda_W,\lambda_t^\prime,\lambda_W^\prime) =\frac{1}{6}\bigg[I_{2\times 2}\otimes I_{3\times 3} + \vec{P}^t\cdot\vec{\sigma}\otimes I_{3\times 3}\notag\\ 
 + \frac{3}{2}I_{2\times 2}\otimes \vec{P}^W\cdot\vec{S} + \sqrt{\frac{3}{2}}I_{2\times 2}\otimes T_{ij}^W\{S_i,S_j\} \notag \\ 
+  pp_{ij}^{tW}\sigma_i\otimes S_j + pT_{ijk}^{tW}\sigma_i\otimes\{S_j,S_k\}\bigg]\notag\\ \left(i,j\in[x\equiv 1,y\equiv 2, z\equiv 3]\right)~
\end{eqnarray}
with $\lambda_{t(W)}$ as the helicity of $t(W)$, $\sigma$ as the Pauli spin matrices, and $S$ as the spin matrices for spin-$1$ particle.
Here, $p_i^{t/W}$ are the vector polarizations of the top quark and $W$ boson, $T_{ij}^W$ are the tensor polarizations of $W$ boson, $pp_{ij}^{tW}$ are the vector-vector spin correlations, and $pT^{tW}_{ijk}$ represent the vector-tensor spin correlations of the top quark and the $W$ boson. 
The tensor $pT^{tW}$ are symmetric and traceless under the last two indices, similar to the case of $T^W$. Therefore, in the $tW$ process, we have $6$ vector and $5$ tensor polarization parameters; $9$ vector-vector spin correlations, and $15$ vector-tensor spin correlations parameters making a total $35$ independent polarization and spin correlation parameters. These polarization and spin correlation parameters can be obtained from the distribution of the top quark the and $W$ boson's decay products obtained at their rest frame. The joint angular distribution of the leptons coming from both top quark and $W$ (in their respective rest frame) is given by~\cite{Rahaman:2021fcz},
\begin{widetext}
        \begin{eqnarray}    \label{eqn::jdm}
            \frac{1}{\sigma}\frac{d\sigma}{d\Omega_{\ell_t} d\Omega_{\ell_W}}
           &=&\frac{1}{16\pi^2}\bigg[1+\alpha_tp^t_ic_i^{\ell_t}+\frac{3}{2}\alpha_Wp_i^Wc_i^{\ell_W}+\sqrt{\frac{3}{2}}(1-3\delta_W)T_{ij}^Wc_i^bc_j^{\ell_W}(i\neq j)  \notag\\
            &+& \frac{1}{2}\sqrt{\frac{3}{2}}(1-3\delta_W)(T^W_{11}-T^W_{22}) 
             \left((c_1^{\ell_W})^2-(c_2^{\ell_W})^2\right) + \frac{1}{2}\sqrt{\frac{3}{2}}(1-3\delta_W)T^W_{33}\left(3(c_3^{\ell_W})^2-1\right) \notag\\
             &+& \alpha_t\alpha_W pp^{tW}_{ij}c_i^{\ell_t}c_j^{\ell_W}  
             + \alpha_t(1-3\delta_W)pT^{tW}_{ijk}c^{\ell_t}_ic_j^{\ell_W}c_k^{\ell_W}(j\neq k)  \notag\\
           &+&  \frac{1}{2}\alpha_W(1-3\delta_W)\left(pT^{tW}_{i11}-pT^{tW}_{i22}\right)c_i^{\ell_t}\left((c_1^{\ell_t})^2-(c_2^{\ell_W})^2\right) 
             + \frac{1}{2}\alpha_t(1-3\delta_W)pT^{tW}_{i33}c_i^{\ell_t}\left(3(c_3^{\ell_W})^2-1\right)\bigg]
        \end{eqnarray}
 
\end{widetext}
with $\sigma$ as the total cross~section including the decay and $d\Omega=\sin\theta d\theta d\phi$ as the measure of the solid angle of the daughter leptons.
Here, $c_x\equiv c_1$, $c_y\equiv c_2$, and $c_z\equiv c_3$ are the angular functions of the leptons, i.e.,
\begin{eqnarray}\label{eq:cor-defination}
c_{x}^\ell = \sin\theta_\ell\cos\phi_\ell,~c_{y}^\ell = \sin\theta_\ell\sin\phi_\ell,~c_{z}^\ell = \cos\theta_\ell 
\end{eqnarray}
and $\ell_{t(W)}$ denotes lepton decayed from the top quark ($W$ boson). The  analyzing powers $\alpha$  have the values $\alpha_t=1=-\alpha_{\bar{t}}$, $\alpha_{W^\pm} = \pm 1$, and $\delta_W=0$ in the SM~\cite{Boudjema:2009fz}. We have neglected any possible anomalous $tbW$ couplings in the top quark decay. 

The polarization and spin correlation parameters can be obtained by constructing some asymmetries partially integrating with respect to $\theta$ and $\phi$ of the decay products. For example, the tensor polarization $T_{12}^W$ of the $W$ boson can be obtained from the asymmetry~\cite{Rahaman:2016pqj},
\begin{equation}\label{eq:asym-T12W}
    \begin{aligned}
        \mathcal{A}[T^W_{12}] &= \left(\int_{\theta_{\ell_W}=0}^\pi\int_{\phi_{\ell_W}=0}^\frac{\pi}{2}-\int_{\theta_{\ell_W}=0}^\pi\int_{\phi_{\ell_W}=\frac{\pi}{2}}^\pi \right.\\&+\left. \int_{\theta_{\ell_W}=0}^\pi\int_{\phi_{\ell_W}=\pi}^\frac{3\pi}{2} -\int_{\theta_{\ell_W}=0}^\pi\int_{\phi_{\ell_W}=\frac{3\pi}{2}}^{2\pi}\right)d\Omega_{\ell_W} \\&\times \left(\frac{1}{\sigma}\frac{d\sigma}{d\Omega_{\ell_W}}\right) \\
        &= \frac{2}{\pi}\sqrt{\frac{2}{3}}(1-3\delta_W)T^W_{12},
    \end{aligned}
\end{equation}
where the angles  of $\ell_t$ are integrated out. The spin correlation parameters are obtained from asymmetries that involve partial integration of the angles related to both $\ell_t$ and $\ell_W$. The asymmetries for the other polarization  parameters are given by~\cite{Rahaman:2016pqj}, 
\begin{eqnarray}
    \mathcal{A}[p^t_i] &=& \frac{1}{2}\alpha_tp^t_i, \notag\\ 
    \mathcal{A}[p_i^W]&=& \frac{3}{4}\alpha_Bp^B_i,\notag\\
 \mathcal{A}[T^W_{ij}]& =& \frac{2}{\pi}\sqrt{\frac{2}{3}}(1-3\delta_W)T^W_{ij}~(i\ne j),\notag\\
\mathcal{A}[T_{11-22}^W] &=& \frac{1}{\pi}\sqrt{\frac{2}{3}}(1-3\delta_W)T^W_{11-22},\notag\\
      \mathcal{A}[T_{33}^W] &=& \frac{3}{8}\sqrt{\frac{2}{3}}(1-3\delta_W)T^W_{33}, 
\end{eqnarray}
where we have used $T^W_{11-22}=T^W_{11}-T^W_{22}$. The asymmetries for the vector-vector and 
vector-tensor spin correlations are given by~\cite{Rahaman:2021fcz},
\begin{eqnarray}
   \mathcal{A}[pp_{ij}^{tW}]&=&\frac{1}{4}\alpha_t\alpha_W pp_{ij}^{tW},\notag\\ 
        \mathcal{A}[pT^{tW}_{i(jk)}] &=& \frac{2}{3\pi}\alpha_t(1-3\delta_W)pT^{tW}_{i(jk)}, (j \neq k),\notag \\
        \mathcal{A}[pT^{tW}_{i(11-22)}] &=& \frac{1}{3\pi}\alpha_t(1-3\delta_W)pT^{tW}_{i(11-22)},\notag \\
        \mathcal{A}[pT^{tW}_{i33}] &=& \frac{3}{16\pi}\alpha_t(1-3\delta_W)pT^{tW}_{i33}.
\end{eqnarray}
In the above equation, we have denoted $pT^{tW}_{i(11-22)}=pT^{tW}_{i11}-pT^{tW}_{i22}$ in the same spirit of $T_{11-22}$.
For numerical calculation,  the asymmetries related to polarization and spin correlation parameters can be obtained as 
\begin{eqnarray}
	\label{eq:workAsymm}
	{\cal A}_i^{\rm Pol} &=& \dfrac{\sigma\left({\cal C}_i^{\ell_t/\ell_W} >0\right)-\sigma\left({\cal C}_i^{\ell_t/\ell_W} <0\right)}
	{\sigma\left({\cal C}_i^{\ell_t/\ell_W} >0\right)+\sigma\left({\cal C}_i^{\ell_t/\ell_W} <0\right)},\notag\\
	\mathcal{A}_{ij}^{\rm Corr} &=& \frac{\sigma({\cal C}^{\ell_t}_i{\cal C}^{\ell_W}_j > 0) - \sigma({\cal C}^{\ell_t}_i{\cal C}^{\ell_W}_j < 0)}{\sigma({\cal C}^{\ell_t}_i{\cal C}^{\ell_W}_j > 0) + \sigma({\cal C}^{\ell_t}_i{\cal C}^{\ell_W}_j < 0)}
\end{eqnarray} 
where angular functions
$${\cal C}_i^{\ell_t/\ell_W}  \in\left[c_x,~c_y,~c_z\right]$$
are associated with vector polarization, while for tensor polarization (available only for $W$) following are the angular functions,
$${\cal C}_i^{\ell_W}  \in\left[c_xc_y,~c_xc_z,~c_yc_z,~c_x^2-c_y^2,\\~|\sqrt{c_x^2+c_y^2}|(4c_z^2-1)\right].$$
In Eq.~(\ref{eq:workAsymm}), `Pol' and `Corr' represent the polarization and spin correlation, respectively.
The full machinery of the polarizations and spin correlations in a system of spin-$1/2$ and spin-$1$ particles can be found in Ref.~\cite{Rahaman:2021fcz}.

In the next section, we discuss the reconstruction of four-momenta of the missing neutrinos in order to obtain the polarization and spin correlation in the leptonic $tW$ production process. 
 \section{Reconstruction of neutrino momenta}
 \label{sec::mt2}
The process of our interest in this article is the single production of a top quark associated with a $W$ boson followed by their leptonic decay in a hadronic collider, which amounts to two invisible neutrinos. The construction of polarization and spin correlation asymmetries of the top quark and $W$ boson that we discussed above require us to see the angular distributions of the daughter leptons at the rest frame of their respective mother particles. This 
requires one to reconstruct the momenta of the two missing neutrinos in our process. We follow the MAOS reconstruction method described in Ref.~\cite{Cho:2008tj}, and for completeness, we describe the procedure here. The reaction of our interest for the $tW^-$ production can be defined with individual momenta as,
\begin{equation}
\begin{aligned}
pp &\to t(1)~W^-(2) \\ &\to b(p_1)~l^+(p_2)~\nu_l(k_1)~l^-(p_3)~\bar{\nu}_l(k_2).
\end{aligned}
\label{process}
\end{equation}
The production of anti-top quark along with $W^+$ boson would be the charge conjugation of the process given in Eq.~(\ref{process}). Assuming that the top quark and the $W$ boson are produced on-shell, we have the following on-shell constraints,
\begin{eqnarray}
\label{eq:cons-onshell}
&	(p_1 + p_2 + k_1)^2 = m_t^2,& \notag\\ 
&(p_3 + k_2)^2 = m_W^2,& \notag\\  
&k_i^2 = 0.&\\ 
\end{eqnarray} 
One more constraint arises due to the total missing transverse momenta given by,
\begin{equation}\label{eq:cons-pmiss}
\vec{\slashed{p}}_T = \vec{k}_{1T} + \vec{k}_{2T},
\end{equation}
where   $\vec{k}_{1T}$ and $\vec{k}_{2T}$ are the transverse momenta of the two neutrinos. 
Equations~(\ref{eq:cons-onshell}) and (\ref{eq:cons-pmiss}) provide $6$ constraints on $8$ unknowns, $k_i^\mu (i = 1,2)$, and we parameterized the $2$-parameter family of solutions by $\vec{k}_{1T}$. For any real choice of $\vec{k}_{1T} = \tilde{k}_{1T}$, the longitudinal components of neutrino momenta are given by,
\begin{align}    	\label{eqn:mt2sols}
k_{iL} = \frac{1}{(E_{iT}^{V})^2} \bigg[p_{iL}A_i \pm \sqrt{p_{iL}^2 + (E_{iT}^{V})^2} \notag\\\times \sqrt{A_i^2 - (E_{iT}^V)^2(E_{iT}^{\chi})^2}\bigg], 
\end{align}
where $V$ and $\chi$ represent visible and invisible (neutrinos) particles, respectively. Here, $E_{iT}$ is the transverse energy and $A_i$ is defined as,
\begin{equation}
A_i = \frac{1}{2}\left(m_W^2-m_\chi^2-p_i^2\right) + \vec{p}_{iT}\cdot\vec{k}_{iT}.
\end{equation}
The solutions for the longitudinal components of the neutrinos will be real if and only if $|A_i| \geq E_{iT}^VE_{iT}^{\chi}$ which is equivalent to
\begin{equation}
\label{condmt2}
m_W \geq \text{max}\{ M_T^{(1)}, M_T^{(2)}\}
\end{equation} 
where 
\begin{equation}\label{eq:MT}
M_T^{(i)} = \sqrt{p_i^2 + 2\left(E_{iT}^VE_{iT}^{\chi} - \vec{p}_{iT}\cdot \vec{k}_{iT}\right)}
\end{equation}
corresponds to the transverse mass of the $W$s and $p_i$s are the four-momenta of the visible daughter of $W$ boson. From Eq.~(\ref{condmt2}), it can be found that the best choice of $\widetilde{\vec{k}}_T$ is the one minimizing $\text{max}\{M_T^{(1)},M_T^{(2)}\}$, i.e., the value giving the collider variable $M_{T2}$~\cite{Lester:1999tx,Barr:2003rg}: 
\begin{equation}
M_{T2}(p_i,m_{\chi}) \equiv 
\mathop{\text{min}}_{\vec{k}_{1T} + \vec{k}_{2T} = \vec{\slashed{p}}}
[\text{max}\{M_T^{(1)},M_T^{(2)}\}]
\end{equation}
where $m_\chi$ is the assumed mass of the invisible particle, which is set to zero. In cases where the solutions are imaginary, the values of $m_W$ are varied $1000$ times following the normal distribution with a mean of $80.4$ GeV and standard deviation of $20$ GeV. Events are only accepted if real solutions are found within the given iteration. 
In general, as can be seen in Eq.~(\ref{eqn:mt2sols}), two solutions exist for the longitudinal momenta of each neutrino. In the case of neutrino from the top quark side, we use the on-shell top mass constraint, i.e., we choose the solution that provides the minimum  $|m_W^{True}-m_W^{Reco}|$. With the chosen neutrino solution coming from the top chain, we then choose the solution of neutrino momenta coming from $W^-$ decay by demanding minimum reconstructed partonic center of mass energy $\sqrt{\hat{s}}\equiv m_{tW^-}$.

\subsection{Reconstruction of angular variables}
\begin{figure*}[!htb]
	\centering
	\subfigure{\includegraphics[scale=0.75]{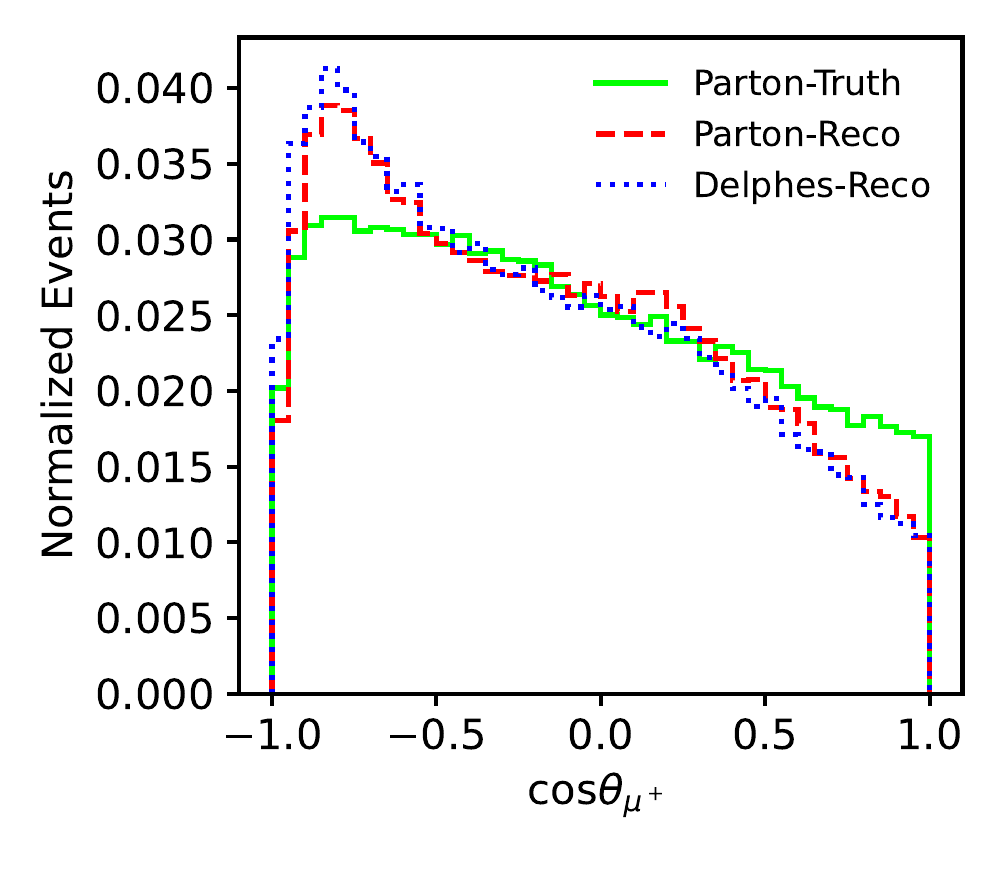}}
	\subfigure{\includegraphics[scale=0.75]{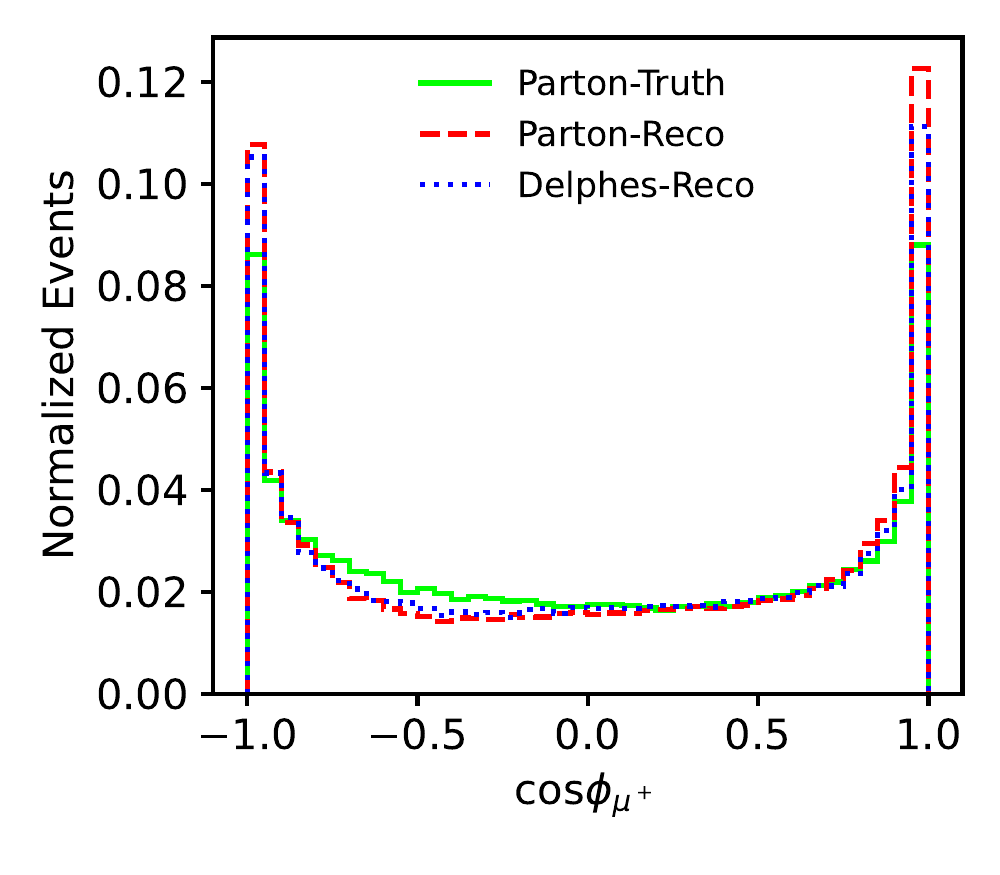}}
	\subfigure{\includegraphics[scale=0.75]{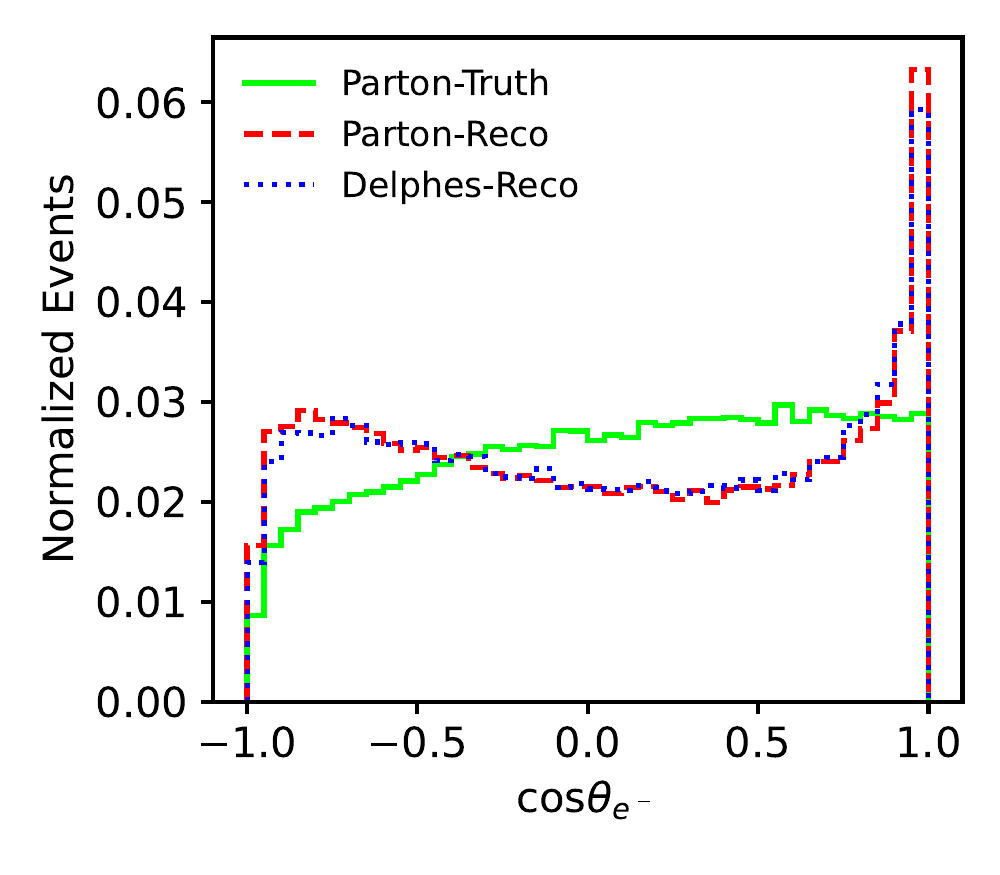}}
	\subfigure{\includegraphics[scale=0.75]{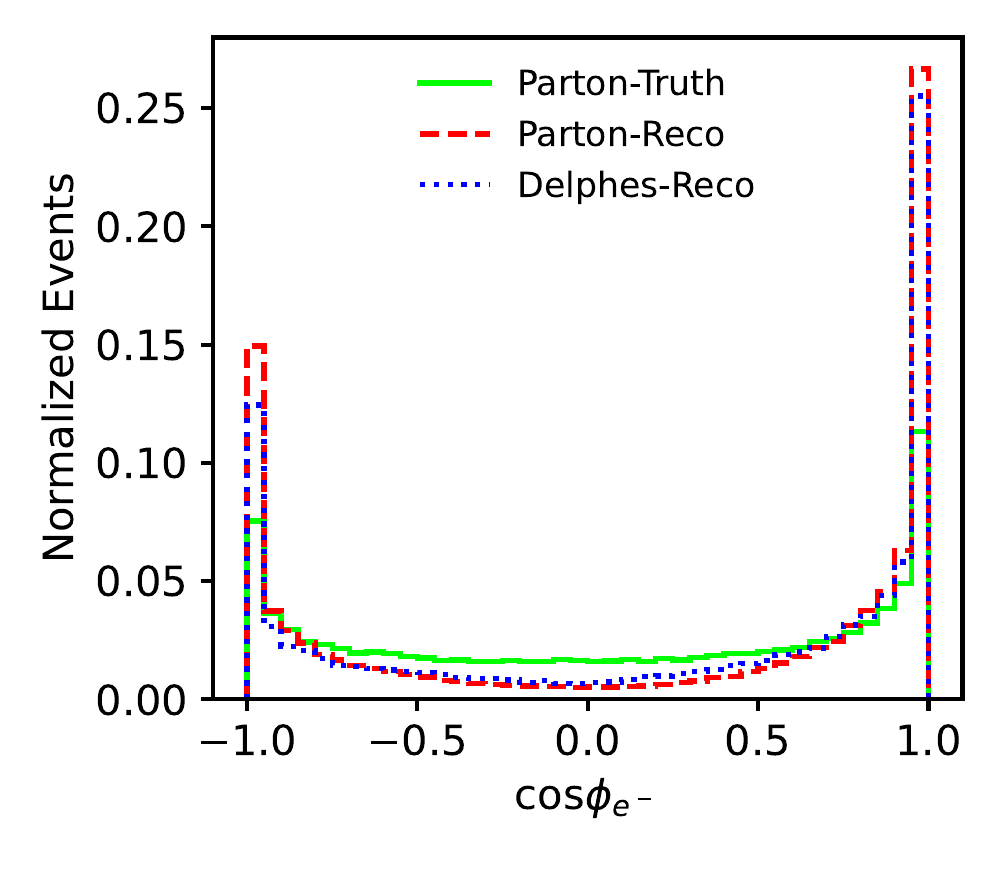}}
	\caption{\label{fig::truth_vs_mt2}Distribution of $\cos\theta$ and $\cos\phi$ of muon coming from top quark ({\em top-row}), and electron coming from $W^-$ ({\em bottom-row}) calculated at the rest frame of the top quark and $W^-$ boson, respectively in the SM for the three cases of Parton-Truth, Parton-Reco, and Delphes-Reco.}
 \end{figure*}
We study the goodness of the MAOS method for the measurement of polarization and spin correlation parameters in the $tW^-$ process by comparing the reconstructed angular distributions with the truth level results. We simulate the events for the $tW^-$ process in the $t\to b\mu^+\nu_\mu,~ W^-\to e^-\bar{\nu_\mu}$ channel using {\tt MadGraph5$\_$aMC$@$NLO~(v2.6.7)}~\cite{Alwall:2011uj} at the $\sqrt{s}=13$ TeV LHC framework using a diagonal CKM matrix at LO in QCD. A flavor factor of $4$ is accounted for in the analysis. We use {\tt NNPDF31}~\cite{NNPDF:2017mvq} set for the parton distribution functions (PDFs) with $\alpha_s(m_Z)=0.118$. We applied some minimum kinematic cuts such as the transverse momenta ($p_T$), pseudorapidity ($\eta$), and geometric distance ($\Delta R = \sqrt{(\Delta\phi)^2 + (\Delta\eta)^2}$)  on the final state jets and leptons while generating events. The generation level cuts are
\begin{eqnarray}\label{eq:gen-cut}
p_T(j) \ge 20~{\rm GeV},~~p_T(\ell) \ge 10~{\rm GeV},\notag\\
|\eta(j)|\le 4.5,~~|\eta(\ell)| \le 2.5,~~\Delta R (\ell/j,\ell) \ge 0.4.
\end{eqnarray}
The generated events are then passed to {\tt Pythia~(v8.306)}~\cite{Sjostrand:2007gs} for showering and hadronization, and further detector response is simulated using the {\tt Delphes~(v3.5.0)}~\cite{deFavereau:2013fsa} package. After the {\tt Delphes} simulation, the objects undergo further selection cuts. The jets are reconstructed with {\tt Anti-$K_T$} algorithm~\cite{Cacciari:2008gp} with isolation radius parameter $R_0(j)=0.5$ with minimum transverse energy of $20$ GeV. The jets are identified as $b$-jet~($b$-tagged) using the $b$-tagging algorithm implemented in {\tt Delphes}. The leptons go through the isolation criteria as follows: The leptons are said to be isolated if the ratio (labeled as isolation variable $I_l$) of the scalar $p_T$ sum of photonic and hadronic activity to the lepton $p_T$ within the isolation cone $R_{\text{max}}(l)=0.5$ is below some small number. For the electron, we have used $I_e < 0.12$, and for the muon, we choose $I_\mu < 0.25$ for its better detectability.    Events are selected at the detector level with exactly one electron, one positively charged muon, and one $b$-tagged jet with the following selection cuts:
	\begin{eqnarray}\label{eq:sel-cut}
		&p_T(b)>20~\text{GeV},~p_T(l)>10~\text{GeV},~|\eta_b|<2.5,\notag\\ &|\eta_l|<2.5.
	\end{eqnarray}
 
The four momenta of these objects and the missing energy are used as input for our MAOS algorithm to reconstruct the momenta of two neutrinos. We compare some angular distributions related to the polarization of $t$ and $W^-$, calculated using parton-level events with Monte Carlo truth information of neutrino momenta (Parton-Truth), with MAOS-reconstructed neutrino momenta (Parton-Reco). We also compare them using the {\tt Delphes} level events with MAOS-reconstructed neutrinos (Delphes-Reco). The normalized distributions of $\cos\theta$ and $\cos\phi$ for the muon coming from the top quark ({\em top-row}) and the electron coming from $W^-$ ({\em bottom-row}), calculated at the rest frame of the top quark and $W^-$, respectively, are shown in Fig.~\ref{fig::truth_vs_mt2} for the three cases of Parton-Truth, Parton-Reco, and Delphes-Reco.

The reconstructed distribution of $\cos\phi$ at the parton and detector level remains close to the true partonic distribution for both the muon and the electron. However, for the distribution of $\cos\theta$, the reconstructed level shows changes compared to the Parton-Truth result, with an overall increase in asymmetry. This increased asymmetry benefits its measurement over the background with low statistics. The reconstructed distributions, both at the parton level and the Delphes level, remain roughly the same.

Although we can reconstruct a few spin angular functions, we cannot reconstruct others at the detector level with the MAOS method. Nevertheless, we calculate the polarization and spin correlation variables at the detector level by reconstructing the missing neutrinos using the MAOS method, and we use them to probe anomalous couplings in the next section.

\section{Probing the anomalous couplings}
\label{sec::probe}
\begin{figure*}[!htb]
	\centering
	\subfigure{\includegraphics[scale=0.75]{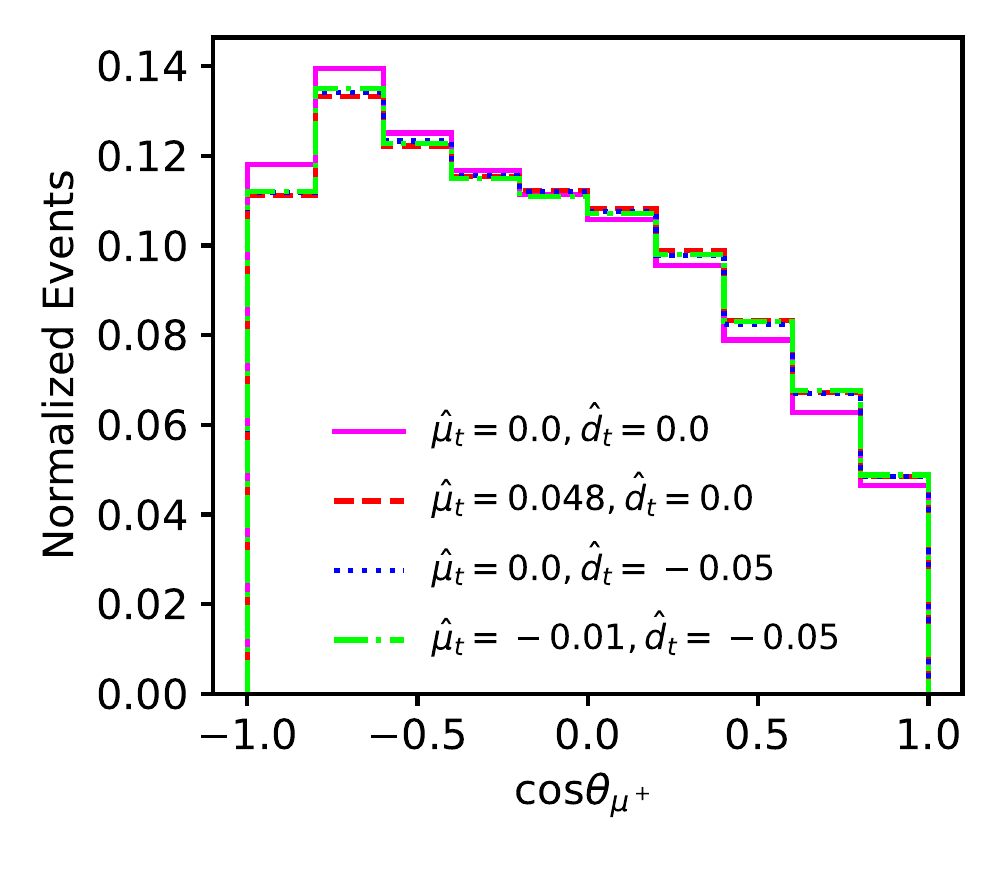}}
        \subfigure{\includegraphics[scale=0.75]{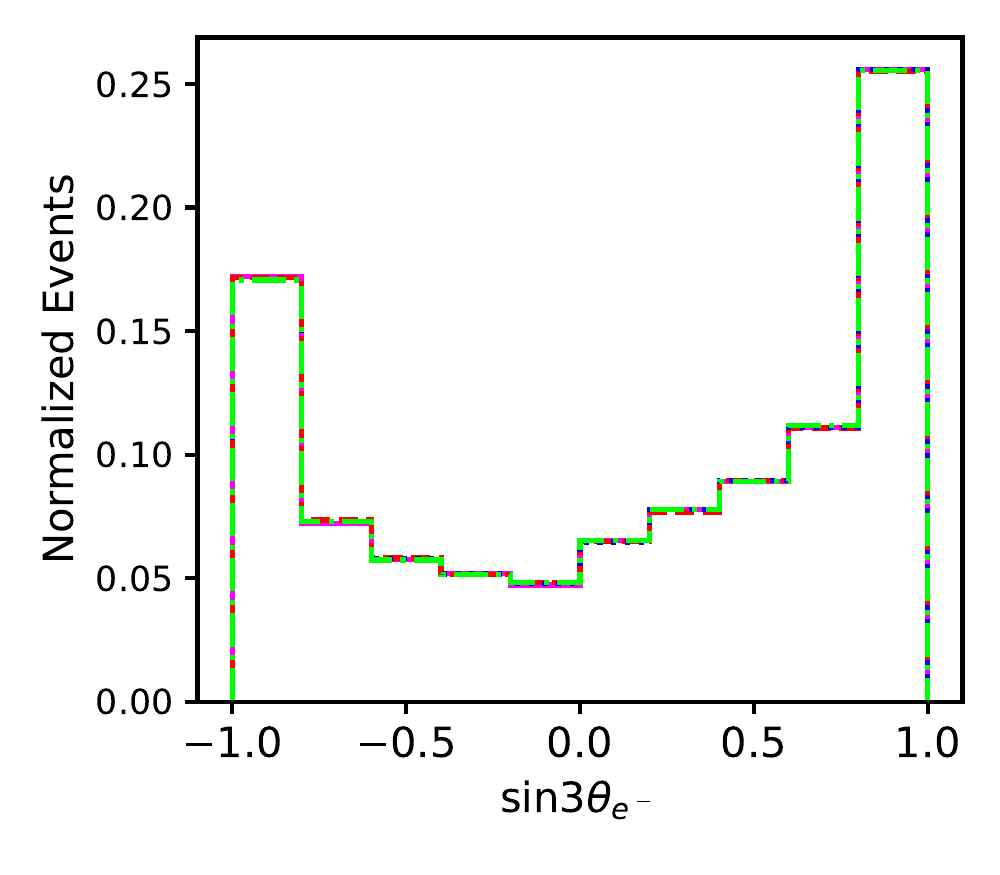}}
        \subfigure{\includegraphics[scale=0.75]{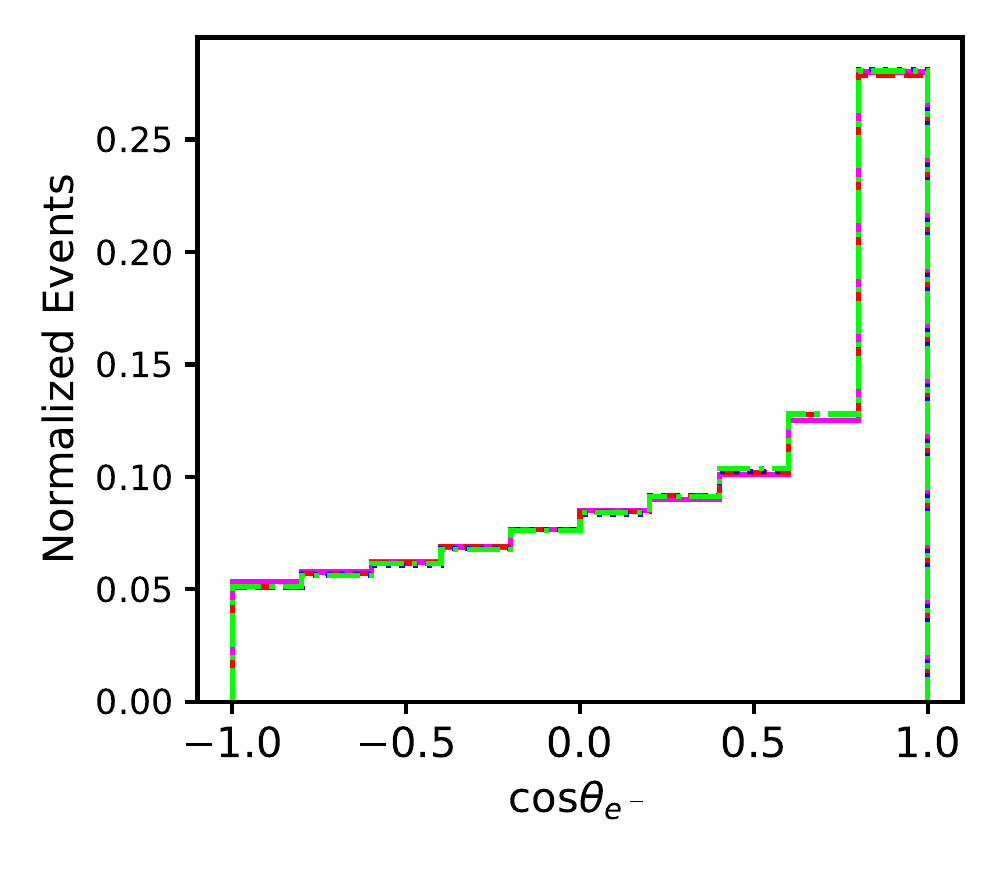}}
	\subfigure{\includegraphics[scale=0.75]{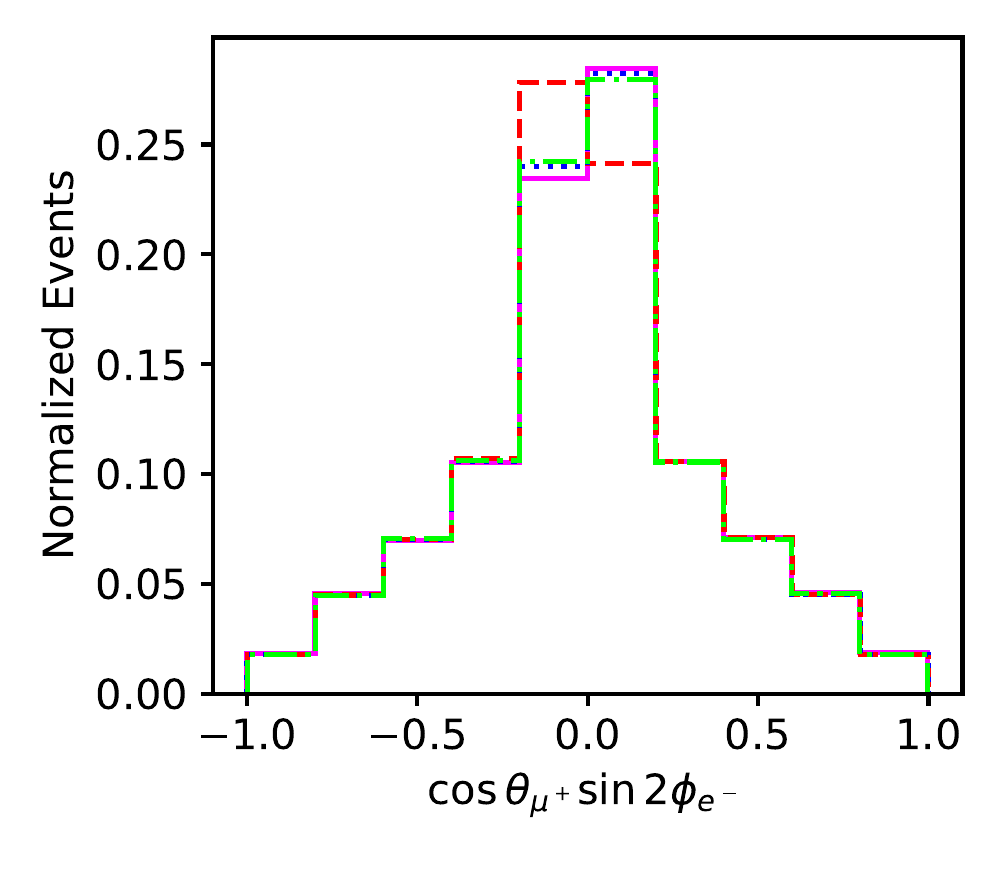}}
	\subfigure{\includegraphics[scale=0.75]{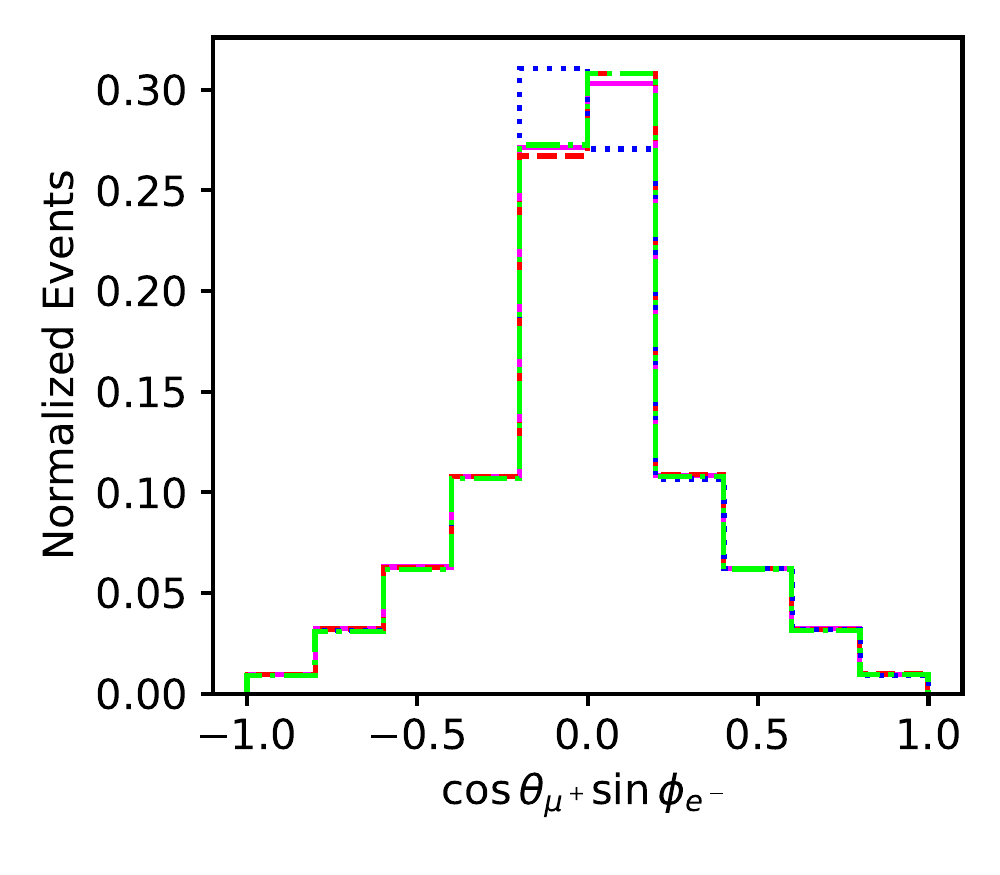}}
	\subfigure{\includegraphics[scale=0.75]{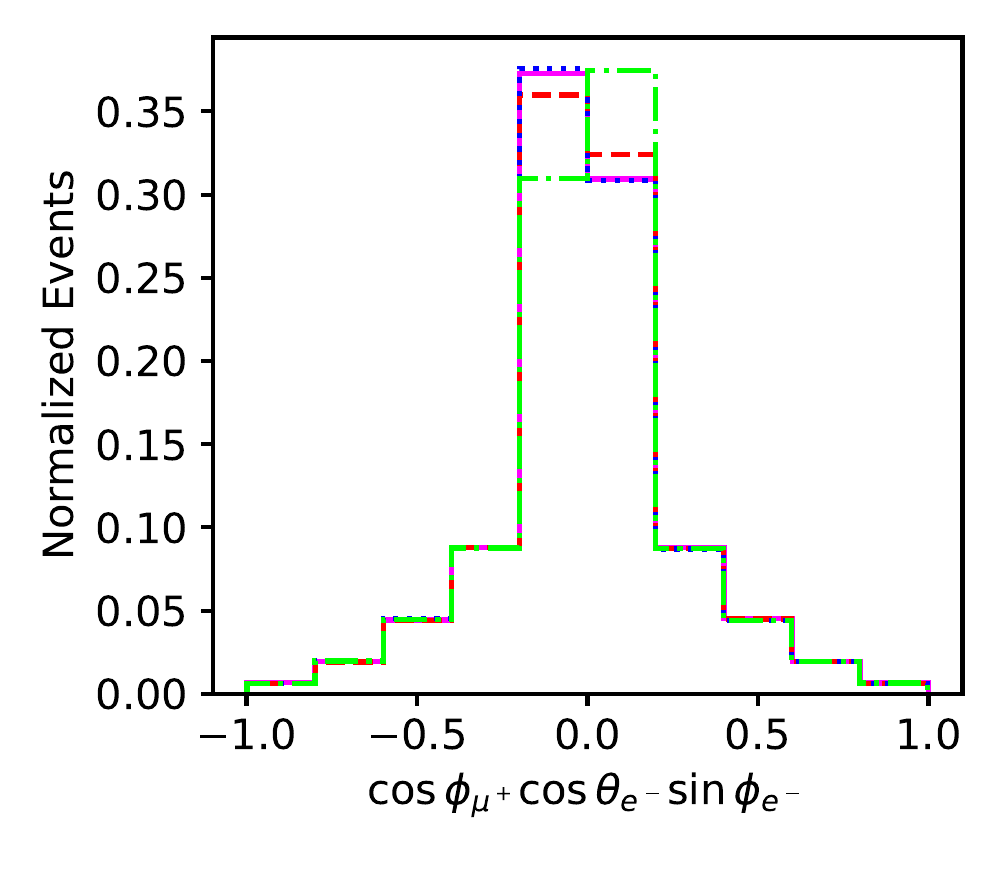}}
	\caption{\label{fig::delphes_sm_vs_eft}Reconstructed detector level distribution of angular functions related to some of the polarization and spin-correlation parameters for the top quark and $W^-$ boson at their respective rest frame.
  The distribution is obtained for SM value and three benchmark points for anomalous couplings ($\hat{\mu}_t, \hat{d}_t$). }
 \end{figure*}

In this section, we study the anomalous couplings $\hat{\mu}_t$ and $\hat{d}_t$ related to top quark chromo-magnetic and chromo-electric moment as introduced in Eq.~(\ref{eq:Laggtt}) in the leptonic $tW$ process using the polarization and spin correlations, along with the total rate. 
 The effect of the anomalous couplings on the matrix element is included through Universal FeynRules Output~(UFO)~\cite{Degrande:2011ua} model file for the Lagrangian given in Eq.~(\ref{eq:Laggtt}) generated in {\tt FeynRules}~\cite{Alloul:2013bka} package. We generated the signal events with anomalous couplings in {\tt Madgraph5} by varying one parameter at a time while keeping the other zero and varying both parameters simultaneously. Parton level events are subsequently passed through {\tt Pythia} and {\tt Delphes} to simulate hadronization and detector effects, respectively. The events are generated in {\tt Madgraph5}  and selected at {\tt Delphes} level with the same set of cuts as given in Eq.~(\ref{eq:gen-cut}) and (\ref{eq:sel-cut}), respectively. 

  The dominant background for the $tW$ process comes from $t\bar{t}$ and $WWj$ process. To keep the backgrounds as kinematically similar to the signal as possible, the selection requirements applied to these backgrounds are identical to those of the signal in the generation level in {\tt Madgraph5} as well as in the detector level in {\tt Delphes}.

The detector efficiency, denoted as $\epsilon$, is determined by calculating the ratio of events that survive the selection criteria to the total number of generated events. The expected number of events from various processes that satisfy all selection cuts can be calculated as follows,
\begin{equation}
	N = \epsilon\times\sigma\times\mathcal{L},
\end{equation}
where $\sigma,~\mathcal{L}$ are the cross~section in each channel and integrated luminosity, respectively. 

To have an idea about the strength of the $tW$($tW^-+\bar{t}W^+$) process in comparison to the backgrounds, we list the value of cross~sections, detection efficiency, and the total number of expected events in Table~\ref{tab:baknum} for a luminosity of $\mathcal{L}=150$~fb$^{-1}$. The number of events is adjusted to account for the QCD NLO effect with the NLO-$k$ factors of $1.34$ for the $tW$ signal~\cite{Kidonakis:2021vob}, $1.47$ for the $t\bar{t}$ background~\cite{Alwall:2014hca}, and $1.3$ for the  $W^-W^+j$ background~\cite{Alwall:2014hca}. We have also taken into account a flavor factor of 4 to incorporate other leptonic decay modes of the top quark and the $W$ boson, in addition to the simulated decays of $t\to\mu$ and $W^\pm\to e^\pm$.
The Table~\ref{tab:baknum} also includes the BSM $tW$ process with the choice of benchmark values for the anomalous couplings of   $(\hat{\mu}_t = 0.048,\hat{d}_t =0)$, $(\hat{\mu}_t =0,\hat{d}_t = -0.05)$and $(\hat{\mu}_t = -0.01,\hat{d}_t = -0.05)$.  The detection efficiency of $0.28$ for the $tW$ signal process is independent of the choice of anomalous couplings. Although the background is huge, the significance ($S/\sqrt{B}$ with $S$ as the signal events and $B$ as the background events) for the SM signal is $42$, which is well above the $5\sigma$ discovery limit. 
\begin{table}
	\centering
	\caption{\label{tab:baknum} cross~section, detection efficiency, and the total number of expected events at a luminosity of $\mathcal{L} = 150$ fb$^{-1}$  for the signal ( fully leptonic $tW$) in the SM and for few choices of anomalous couplings as well as for background are shown
 after passing through {\tt Delphes} selection cuts given in Eq.~(\ref{eq:sel-cut}). The expected numbers of events are adjusted to account for the QCD NLO effect by using the NLO-$k$ factor discussed in texts.}
	\centering
		\renewcommand{\arraystretch}{1.5}
		\begin{tabular*}{0.45\textwidth}{@{\extracolsep{\fill}}lllc@{}}\hline	 Sample&$\sigma$~(pb)&$\epsilon$&Event~($10^3$)\\ \hline
  $tW$(SM)&0.2211&0.28&24.887\\
	 $tW(\hat{d}_t=0.048)$&0.2345&0.28&26.395\\
	 $tW(\hat{\mu}_t=-0.05)$&0.2105&0.28&23.693\\
	 $tW(\hat{d}_t=0.01,\hat{\mu}_t=0.05)$&0.2422&0.28&27.262\\
	 $t\bar{t}$&4.381&0.22&277.99\\
	 $WWj$&0.2796&0.026&2.8351\\ \hline
	\end{tabular*}
\end{table}

We now investigate the effect of anomalous couplings on the polarization and spin correlation objects. 
 Figure~\ref{fig::delphes_sm_vs_eft} depicts the normalized distribution at the detector level for a few angular functions of the leptons originating from the top quark and the $W^-$ boson for their polarizations and spin correlations as representatives in the $tW$ process in SM and the same benchmark anomalous couplings chosen in Table~\ref{tab:baknum}.  
Regarding the polarizations, the anomalous couplings reduce the overall asymmetry in $\cos\theta_{\mu^+}$ ($p_z^t$) as shown in {\em left-top} panel by increasing the number of events for $\cos\theta_{\mu^+}>0$ while reducing in $\cos\theta_{\mu^+}<0$. Meanwhile, for the polarization of $W^-$, $\sin 3\theta_{e^-}$ ($T_{zz}^{W^-}$) and $\cos\theta_{e^-}$ ($p_z^{W^-}$) shown in {\em top-right} and {\em middle-left} panel, respectively remain insensitive to the presence of chosen anomalous couplings. On the other hand, the spin correlation angular functions can differentiate between the $CP$ nature of the anomalous couplings. For example, the correlation angular function $\cos\theta_{\mu^+}\sin2\phi_{e^-}$ ($pT_{z(xy}^{tW^-}$) shown in {\em middle-right} panel changes only for the $CP$-even coupling $\hat{\mu_t}$ by making an overall negative asymmetry. While the correlation $\cos\theta_{\mu^+}\sin\phi_{e^-}$ ($pp_{zy}^{tW^-}$) shown in {\em bottom-left} panel sees the presence of only the $CP$-odd coupling $\hat{d_t}$ by making an overall negative asymmetry.
Thus, by measuring these two spin correlation asymmetries simultaneously, we can probe the $CP$ nature of anomalous couplings if present in the $tW$ process. Additionally, there is one correlation parameter $\cos\phi_{\mu^+}\cos\theta_{e^-}\sin\phi_{e^-}$ ($pT_{xzy}^{tW^-}$) shown in {\em bottom-right} panel which does not independently sense the presence of anomalous couplings but rather their combined effect, i.e., their interference effect.

We find several angular functions related to polarization and correlations, such as $\cos\phi_{\mu^+}$, $\sin2\phi_{e^-}$, and $\sin\phi_{\mu^+}\cos2\phi_{e^-}$, to exhibit symmetry with respect to zero while being less sensitive to the presence of anomalous couplings. For these variables ($x$), we redefine our asymmetries to be,
\begin{equation}\label{eq:mod-asym}
\tilde{A} = \frac{\sigma(|x|>0.5)-\sigma(|x|<0.5)}{\sigma(|x|>0.5)+\sigma(|x|<0.5)}.
\end{equation}
This process enhances the asymmetries and reduces the statistical error related to the asymmetries. This enhancement of asymmetry increases its sensitivity to anomalous couplings. 
\begin{figure}[!htb]
	\centering
	\includegraphics[width=0.5\textwidth]{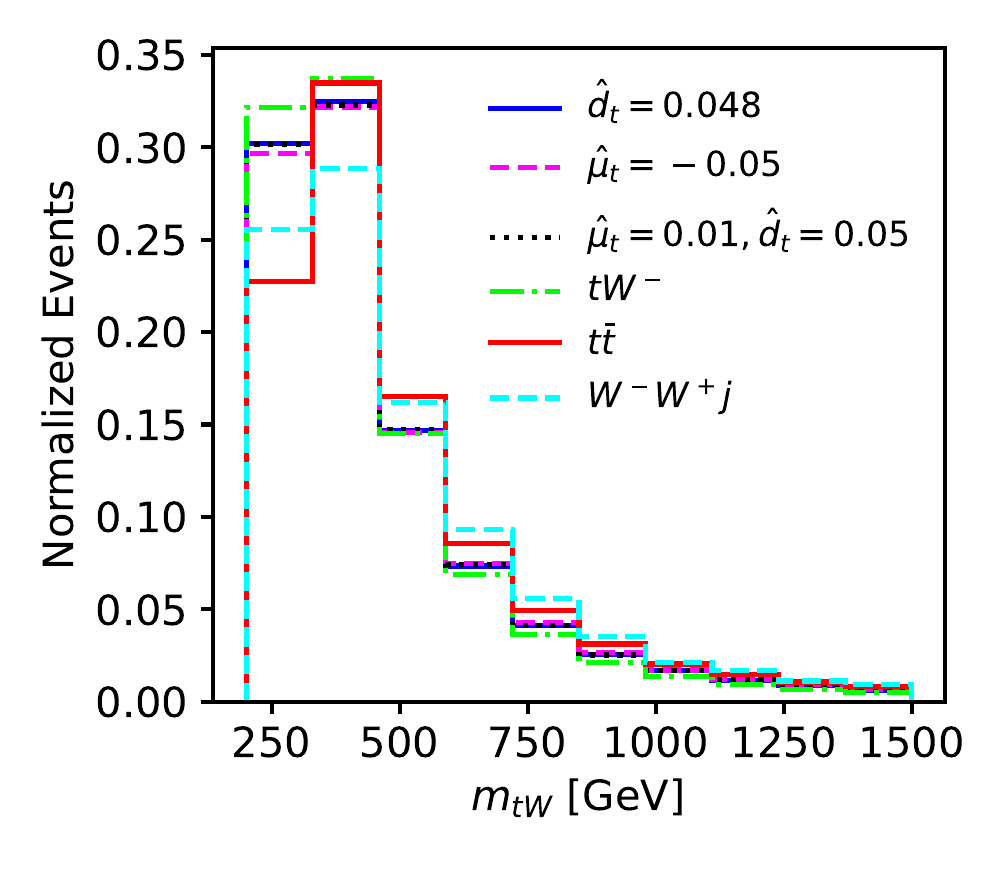}
	\caption{\label{fig::kinematic}Reconstructed detector level invariant mass distribution of top quark and $W^-$ boson~($m_{tW}$) for signal and background processes.}
\end{figure}

In addition to studying the spin angular functions, we investigate several other kinematic distributions to examine the impact of anomalous couplings. These distributions include the reconstructed invariant mass of the top quark and the $W$ boson ($m_{tW}$), the transverse momentum ($p_T$) of the top quark and the $W$ boson, and the missing transverse energy ($\cancel{E}_T$).
Figure~\ref{fig::kinematic} illustrates the normalized distribution of $m_{tW}$, representing the SM signal and chosen benchmark anomalous couplings and background contributions. The vertical axis represents the heights of different bins, which vary both below and above the SM rate. Notably, we observe significant effects of anomalous couplings at lower $m_{tW}$ values ($<400$ GeV) rather than at higher values, contrary to the behavior of anomalous couplings associated with a momentum transfer, e.g., in $t\bar{t}Z$ interactions~\cite{Rahaman:2022dwp}.
Specifically, distributions with all anomalous couplings exhibit smaller heights than the SM signal for $m_{tW}<500$ GeV, while the backgrounds are even smaller. Conversely, the anomalous distributions demonstrate higher heights than the SM signal for $m_{tW}>500$ GeV. Exploiting this binned dependence of anomalous couplings, we maximize the outcome of the $m_{tW}$ variable by dividing the variable into three regions: $m_{tW}<500$~GeV, $500<m_{tW}<1000$~GeV, and $m_{tW}>1000$~GeV. This binning strategy significantly improves the limits on the anomalous couplings compared to the un-binned case, as discussed in appendix~\ref{Appendix1} using the cross~section alone.

In addition to the cross~section, we compute all the 11 polarization and the 24 spin correlation asymmetries within each bin, thereby obtaining 108 observables to constrain the anomalous couplings effectively. The increase of observables serves our objective of improving the accuracy in constraining anomalous parameters.

We compute all observables in each bin for the SM signal and several sets of benchmark anomalous couplings. We analyze six points for one-parameter variations, and for simultaneous non-zero values of both couplings, we analyze twelve points. These observable values are utilized for numerical fitting to obtain semi-analytical expressions, which will be employed in our analysis.
For the cross~section, which is a $CP$-even observable, we employ the following parametric function to fit the simulated data in the presence of anomalous couplings ($\hat{\mu}_t,\hat{d}_t$):
\begin{equation}
\label{eqn::cpfit}
    \sigma\left(\{\hat{\mu}_t,\hat{d}_t\}\right) = \sigma_0 + \sigma^{(1)}_{\hat{\mu}}\hat{\mu}_t +\sigma^{(2)}_{\hat{\mu}}\hat{\mu}_t^2 + \sigma_{\hat{d}}\hat{d}_t^2.
\end{equation}
Regarding the asymmetries, the denominator is the cross~section, and the numerator represents the difference between two cross~sections, for which we derive the corresponding expressions to calculate the asymmetries. The numerators ($\Delta\sigma = {\cal A}\times\sigma$) of the $CP$-odd asymmetries are fitted using the function:
\begin{equation}
    \Delta\sigma\left(\{\hat{\mu}_t,\hat{d}_t\}\right) = \sigma_{\hat{d}}\hat{d}_t + \sigma_{\hat{\mu}\hat{d}}\hat{\mu}_t\hat{d}_t.
\end{equation}
For the  $CP$-even asymmetries, the numerators are fitted using Eq.~(\ref{eqn::cpfit}).

The non-signal type backgrounds $t\bar{t}$ and $W^+W^-j$ modify the asymmetries. The net asymmetries combining the signal and the backgrounds are given by
\begin{equation}
{\cal A} = \frac{\sum_i {\cal A}_i \sigma_i }{\sum_i\sigma_i},
\end{equation}
where the index $i$ represents the signal and the backgrounds.

\begin{figure*}[!t]
	\centering
	\subfigure{\includegraphics[width=0.46\textwidth]{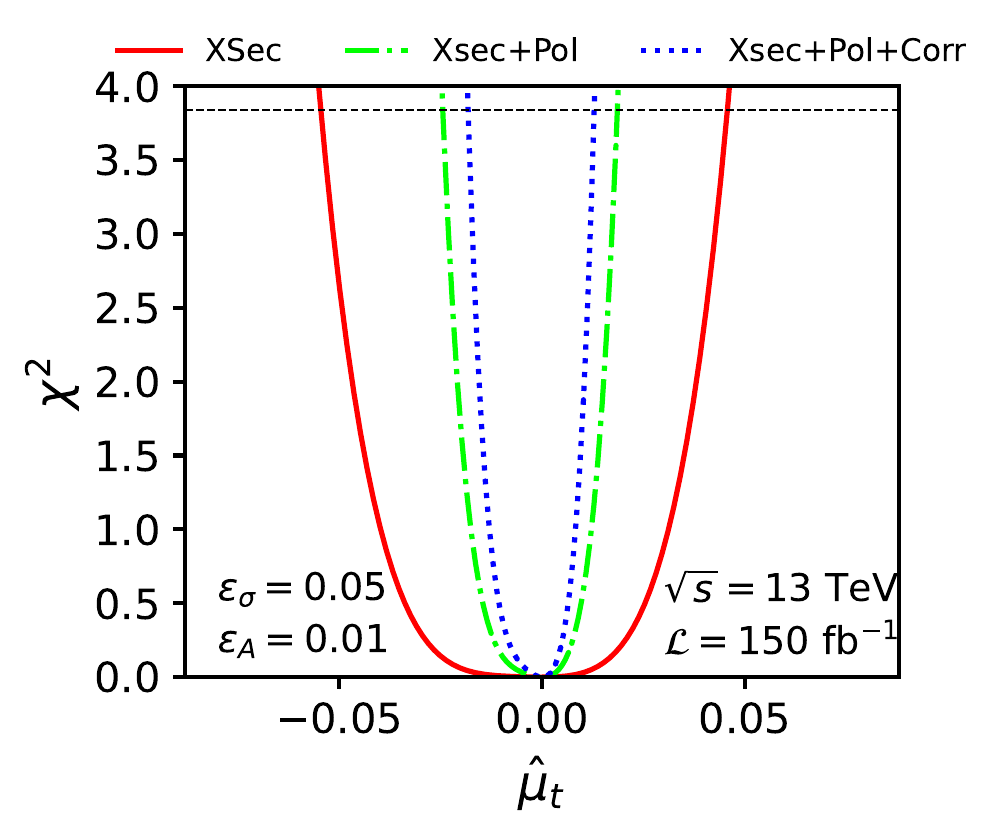}}
	\subfigure{\includegraphics[width=0.46\textwidth]{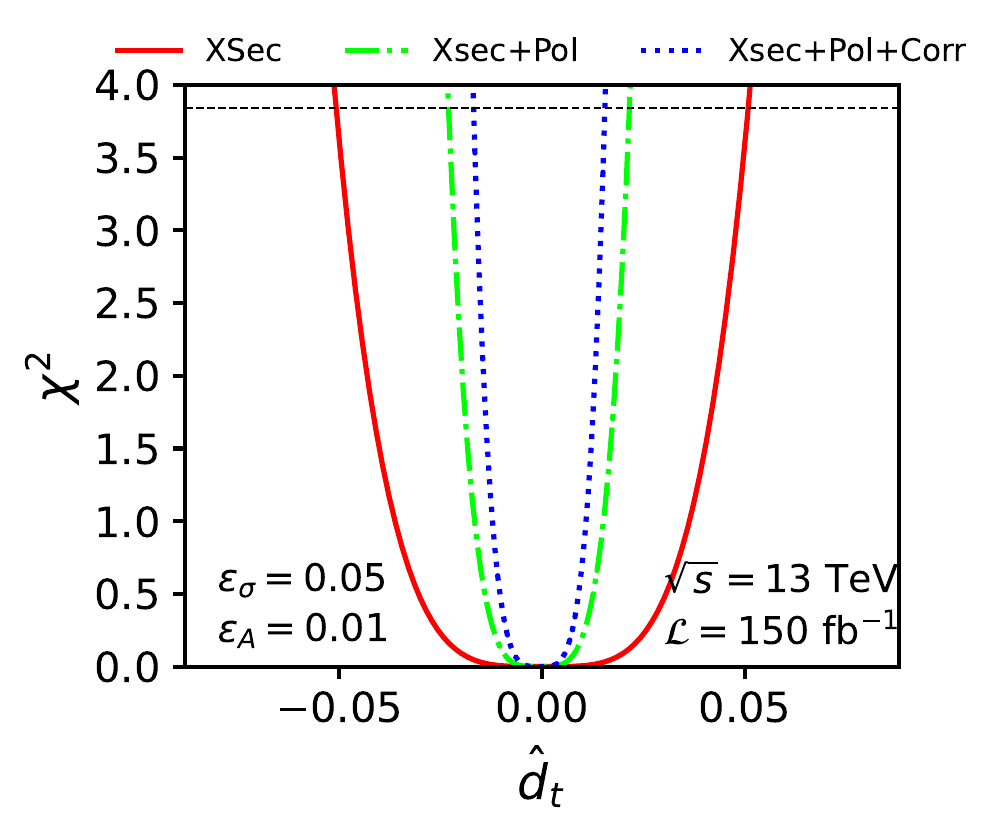}}
	\caption{\label{fig::one_para} One dimensional $\chi^2$ function for cross~sections (XSec) and XSec added with polarizations (Pol) and spin correlations (Corr) successively in all three bins of $m_{tW}$ as a function of anomalous moments  $\hat{\mu}_t$ and $\hat{d}_t$. The $\chi^2$ functions are computed for $\sqrt{s} = 13$ TeV, $\mathcal{L}=150$ fb$^{-1}$ and systematic errors   $\epsilon_\sigma=0.02,\epsilon_A=0.005$.  The horizontal line at $\chi^2=3.84$ represent limit on anomalous couplings at $95\%$ C.L.}
\end{figure*}
With the semi-analytical expressions for the observables in hand, we study the sensitivity of the cross~section, polarization, and spin correlation asymmetries by constructing the $\chi^2$ function of the anomalous couplings. 
Given an observable $\mathscr{O}$, we find the chi-squared distance between an anomalous coupling point and the SM point defined as,
\begin{equation}
	\label{eqn:chisq}
	\chi^2(\vec{x}) = \sum_i\frac{(\mathscr{O}_i(\vec{x}) - \mathscr{O}_i(\vec{x}=0))^2}{(\delta\mathscr{O}_i(\vec{x}=0))^2},
\end{equation}
where $\mathscr{O}_i(\vec{x})$ and $\mathscr{O}_i(0)$ is an observable at anomalous point~($\vec{x}$) and SM value respectively. The index $i$ runs on all the observables in all bins. The quantity $\delta\mathscr{O}(0)$ is the estimated net error in $\mathscr{O}$. The estimated error $\delta\mathscr{O}$ is the quadratic sum of statistical and systematic error, i.e.,
\begin{equation}
	\delta\mathscr{O} = \sqrt{(\delta\mathscr{O}_{\text{stat.}})^2 + (\delta\mathscr{O}_{\text{syst.}})^2}.
\end{equation} 
For  cross~section $\sigma$, the estimated error is defined as:
\begin{equation}
	\delta\sigma = \sqrt{\frac{\sigma}{\mathcal{L}} + (\epsilon_\sigma\sigma)^2},
\end{equation}
and for  asymmetries, it is given by:
\begin{equation}
	\delta A = \sqrt{\frac{1-A^2}{\mathcal{L}\sigma} + \epsilon_A^2},
\end{equation}
where $\sigma,\mathcal{L}$ are the SM cross~section and integrated luminosity, respectively;  $\epsilon_\sigma$ and $\epsilon_A$ are the systematic error in cross~section and asymmetries, respectively. Some of the sources of systematic error are $b$-tagging efficiency, luminosity, parton distribution functions~(PDF), jet energy scale, lepton energy scale, initial state radiation~(ISR), final state radiation~(FSR), etc. 

We studied the sensitivity of the observables by successively adding the polarization and spin correlation asymmetries to the cross~sections in terms of the $\chi^2$ defined in Eq.~(\ref{eqn:chisq}). These sensitivities are shown in Fig.~\ref{fig::one_para} by varying one parameter at a time while setting the other to its SM value, i.e., zero. The sensitivities were computed for an integrated luminosity of $\mathcal{L}=150$ fb$^{-1}$ and benchmark systematic errors $(\epsilon_\sigma,\epsilon_\mathcal{A}) = (0.05,0.01)$. The dashed horizontal line in Fig.~\ref{fig::one_para} corresponds to $\chi^2 = 3.84$, which translates to a $95\%$ confidence level (C.L.)~\cite{Cowan:2010js}. 
The sensitivities for both couplings improve significantly when polarization asymmetries (Pol) are included in the cross~section (XSec). The sensitivities further improve when the spin correlation asymmetries (Corr) come into play. 
The final constraints on the anomalous couplings are determined by combining all the observables, i.e., by combining XSec+Pol+Corr. We note that we have simulated only the process $tW^-$ and included the process $\bar{t}W^+$ in our analysis by accounting for a factor of 2 in the total $\chi^2$. This adjustment is possible because the production cross~section is the same for both processes, and the sensitivities of the asymmetries to the anomalous couplings remain consistent.

We now compare our limits on the anomalous couplings with the existing stringent limits from the experiment~\cite{CMS:2019kzp}, as discussed in section~\ref{sec:intro}.
The experimental investigation focuses on $t\bar{t}$ production at a center-of-mass energy of $\sqrt{s}=13$ TeV, with an integrated luminosity of $\mathcal{L}=35.9$ fb$^{-1}$. We note that in the experimental study, $tW$ serves as a background process, whereas in our analysis, $tW$ is considered the signal process, while $t\bar{t}$ acts as one of the backgrounds. Since the cross~section of $tW$ is much lower compared to $t\bar{t}$, the limit in our case is expected to be looser compared to the experimental limit.
For an integrated luminosity of $\mathcal{L}=35.9$ fb$^{-1}$, we find the limits on the anomalous couplings to be $-0.0189 < \hat{\mu}_t < +0.0133$ and $-0.0175 < \hat{d}_t < +0.0161$, incorporating systematic errors of $(\epsilon_\sigma,\epsilon_A) =(0.05,0.01)$. It is worth noting that our limit for $\hat{d}_t$ is tighter by roughly a factor of $2$ compared to the experimental value. In the case of $\hat{\mu}_t$, though our analysis provides a tighter lower limit, it provides a relatively looser upper bound than the experimental constraints. However, the range of the limit on $\hat{\mu}_t$ from our analysis ($0.0322$) is tighter compared to the experimental one ($0.049$).
\begin{figure}[!h]
	\centering
	\includegraphics[width=0.47\textwidth]{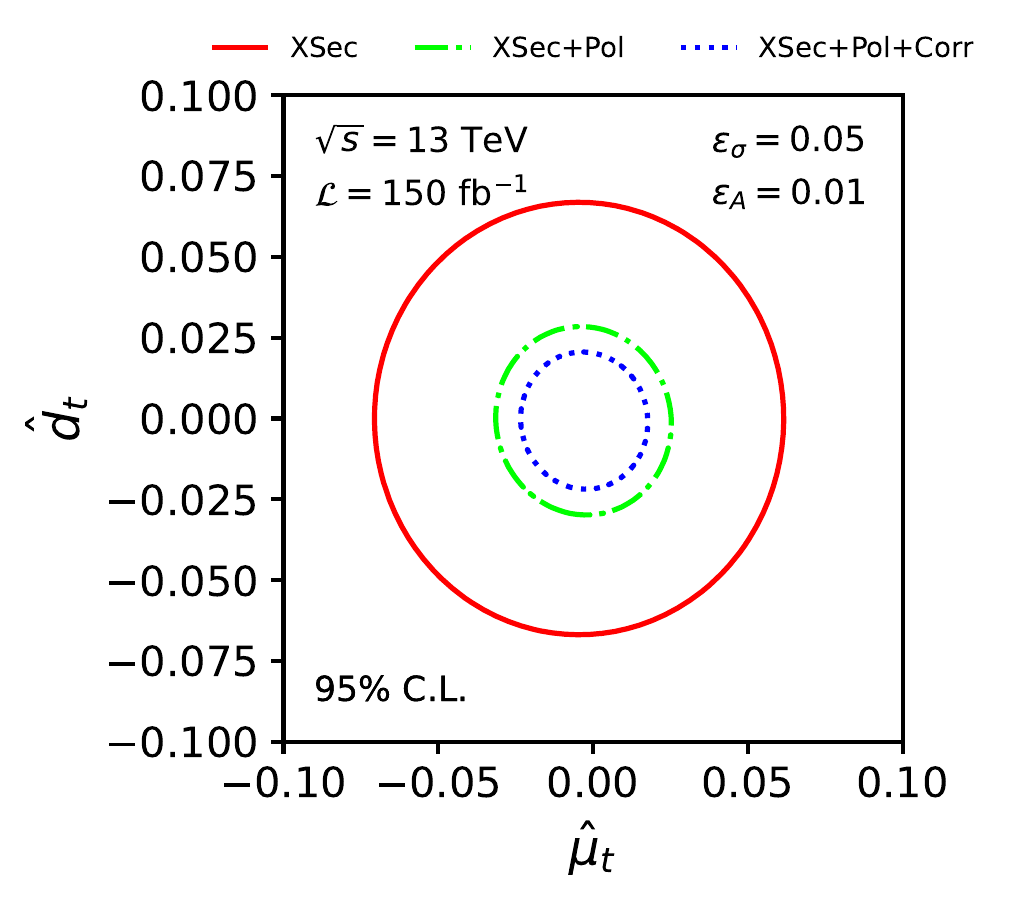}
	\caption{\label{fig::two_para}
 Two dimensional $95\%$ C.L. contours computed with the $\chi^2$ function for cross~sections (XSec) and XSec added with polarizations (Pol) and spin correlations (Corr) successively in all three bins of $m_{tW}$ as a function of anomalous moments  $\hat{\mu}_t$ and $\hat{d}_t$. The contours are obtained for $\sqrt{s} = 13$ TeV, $\mathcal{L}=150$ fb$^{-1}$ and systematic errors   $\epsilon_\sigma=0.02,\epsilon_A=0.005$. }
\end{figure}
In addition to studying the one-parameter sensitivity, we also investigate the two-parameter sensitivity and compare the cross~sections and cross~sections with polarization and correlation asymmetries in terms of $\chi^2$. Figure~\ref{fig::two_para} depicts the $95\%$ confidence level contours ($\chi^2=5.99$~\cite{Cowan:2010js}) with both parameters varying simultaneously, using the same integrated luminosity ($\mathcal{L}=150$ fb$^{-1}$) and systematic errors ($(\epsilon_\sigma,\epsilon_a) = (5\%,1\%)$) as has been considered for the one-parameter case in Fig.~\ref{fig::one_para}. Similar to the one-parameter analysis, the inclusion of polarizations significantly contributes to constraining the anomalous couplings. The addition of spin correlation further tightens the contours when incorporated into the set of observables. It is worth noting that the two anomalous couplings, $\hat{\mu}_t$ and $\hat{d}_t$, remain uncorrelated.

\begin{figure}[!h]
 	\centering
 	\includegraphics[width=0.49\textwidth]{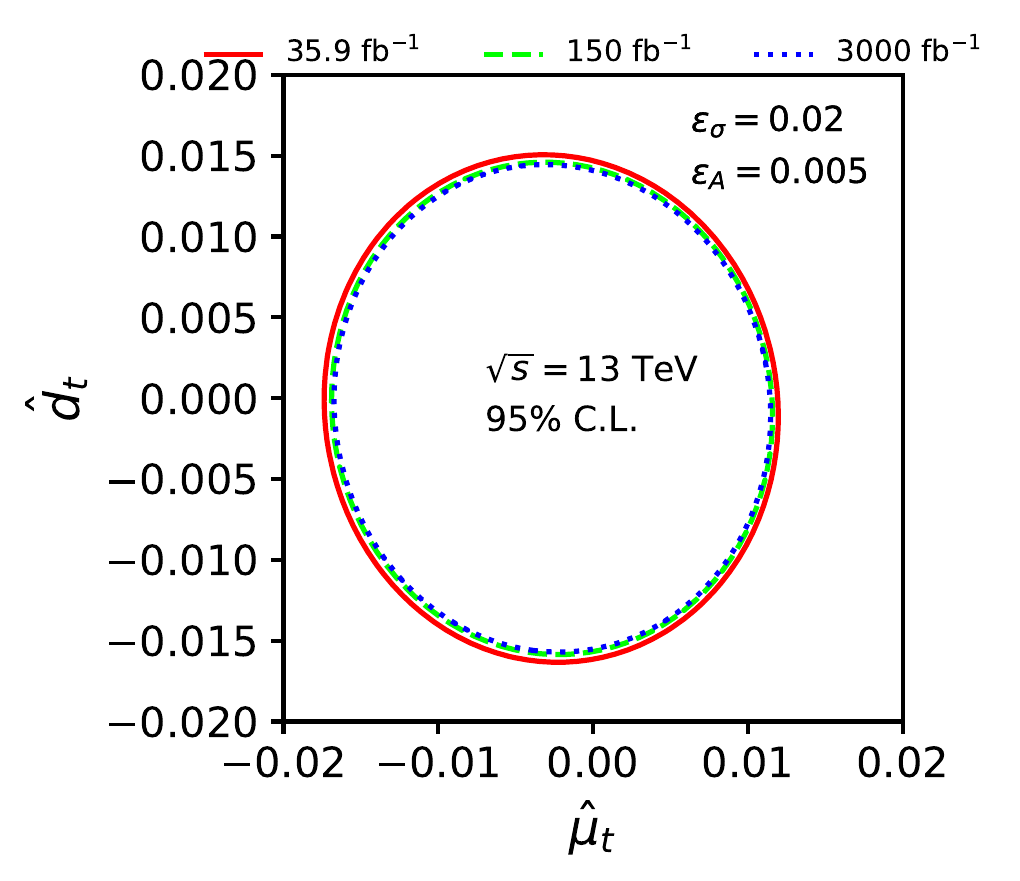}
        \caption{\label{fig::lumisyst} Two dimensional contours at $95\%$ C.L. obtained for the combination of cross~section and all asymmetries as a function of two anomalous couplings for three choices of luminosities and fixed systematic error $(\epsilon_\sigma,\epsilon_A) = (0.02,0.005)$ at $\sqrt{s}=13$~TeV.}
 \end{figure}
 \begin{figure}[!h]
 	\centering
 	\includegraphics[width=0.49\textwidth]{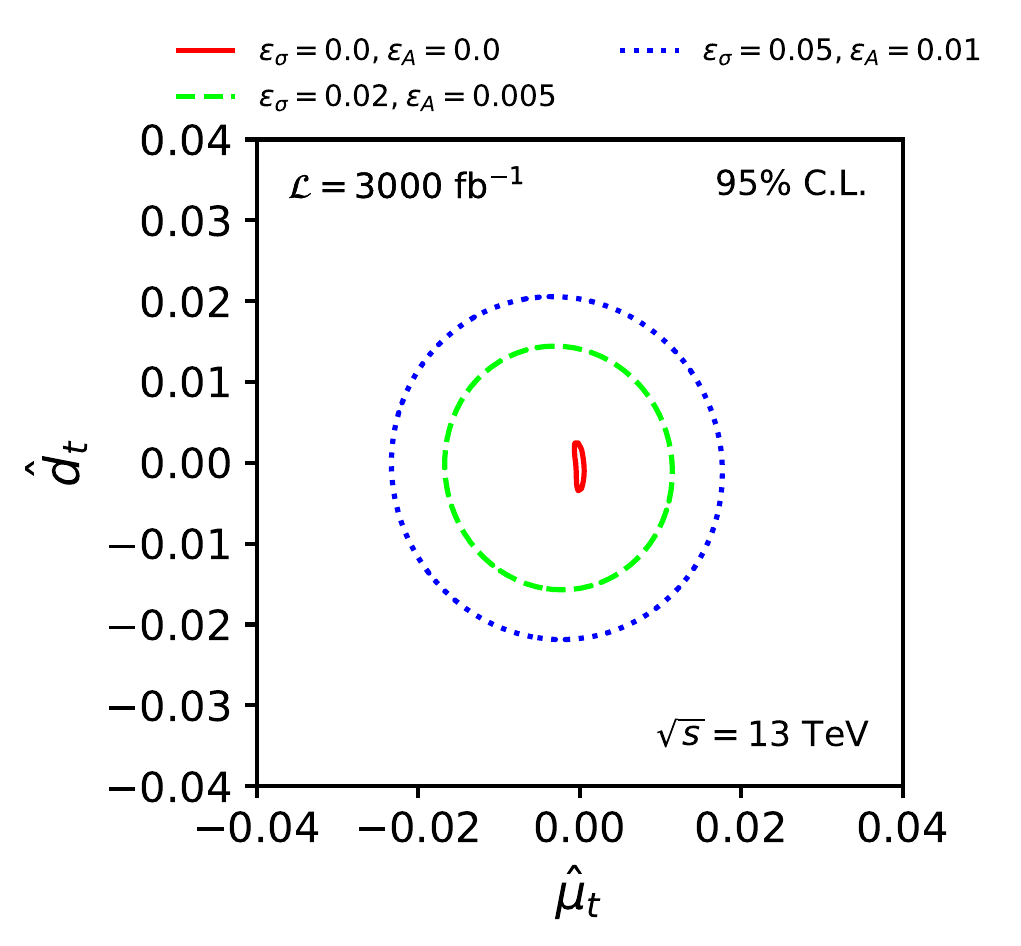}
        \caption{\label{fig::systlumi}Two dimensional contours at $95\%$ C.L. for combinations of cross~section and all asymmetries as a function of two anomalous couplings. The contours are obtained for different sets of systematic errors given in Eq.~(\ref{eqn:syst}), $\mathcal{L} = 3000$ fb$^{-1}$ and center-of-mass energy of 13~TeV.}
 \end{figure}
 To understand how the simultaneous limits on $\hat{\mu}_t$ and $\hat{d}_t$ change with increasing luminosity, we present the two-dimensional $95\%$ C.L. contours in Fig.~\ref{fig::lumisyst} for luminosities of $\mathcal{L} = 35.9$ fb$^{-1}$, $150$ fb$^{-1}$, and $3000$ fb$^{-1}$, with a relatively lower systematic error of ($\epsilon_\sigma,\epsilon_A)=(2\%,0.5\%)$. The contours do not tighten significantly as the luminosity increases from $35.9$ fb$^{-1}$ to $3000$ fb$^{-1}$. This minimal change in the limits of anomalous couplings, even with an 84-fold increase in luminosity, is due to the dominant influence of systematic error, which surpasses the statistical error for higher luminosity. Furthermore, we investigate the role of systematic error on the $95\%$ C.L. contours of the anomalous couplings for a set of systematic errors: $(0.0,0.0)$, $(2\%,0.5\%)$, and $(5\%,1\%)$, with a fixed luminosity of $\mathcal{L}=3000$ fb$^{-1}$, as shown in Fig.~\ref{fig::systlumi}. The contours progressively expand as the systematic errors increase. The radius of the contours increases by approximately a factor of 2 when comparing the systematic errors of $(2\%,0.5\%)$ to $(5\%,1\%)$. 

As a final result, we estimate  the limits on the couplings $\hat{\mu}_t$ and $\hat{d}_t$ and list them in Table~\ref{Tab:final} for a set of integrated luminosities and systematic error given by, 
\begin{equation}
    \label{eqn:lumi}
    \mathcal{L} = \{35.9~\text{fb}^{-1}, 150~\text{fb}^{-1}, 3000~\text{fb}^{-1}\},
\end{equation} 
and 
\begin{equation}	 \label{eqn:syst}
	\begin{aligned}
		\epsilon_\sigma &= \{0.0,0.02,0.05\}, \\
		\epsilon_A &= \{0.0,0.005,0.01\}.
	\end{aligned}
\end{equation}
The limits on the anomalous couplings  are translated to the Wilson's Coefficient of the dimension-$6$ operator given in Eq.~(\ref{eq:operator}) using Eq.~(\ref{eq:lag-vs-eft}) and they are listed in Table~\ref{Tab:final} in last two columns.
 \begin{table*}[]
	\centering
	\renewcommand{\arraystretch}{1.3} 
	\caption{\label{Tab:final} One parameter 95$\%$ C.L. limits on the anomalous chromo-magnetic~($\hat{\mu}_t$) and chromo-electric~($\hat{d}_t$) dipole moments obtained with different values of integrated luminosity and systematic error as given in Eq.~(\ref{eqn:syst}). The translated limits to the effective Wilson's coefficient~(TeV$^{-2}$) are also presented using Eq.~(\ref{eq:lag-vs-eft}).}
	\begin{tabular*}{1\textwidth}{@{\extracolsep{\fill}}cccccc@{}}\hline
		$\mathcal{L}$ (fb$^{-1}$)&($\epsilon_\sigma,\epsilon_{\mathcal{A}}$)& $\hat{\mu}_t$ & $\hat{d}_t$ & $\frac{\text{Re}(C_{uG\Phi})}{\Lambda^2}$~(TeV$^{-2}$) & $\frac{\text{Im}(C_{uG\Phi})}{\Lambda^2}$~(TeV$^{-2}$)\\ \hline			&$(0.0,0.0)$&$[-0.0080,+0.0055]$&$[-0.0112,+0.0097]$&$[-45.990,+31.618]$&$[-64.386,+55.763]$ \\
            $35.9$&$(0.02,0.005)$&$[-0.0147,+0.0095]$&$[-0.0140,+0.0125]$&$[-84.507,+54.613]$&$[-80.483,+71.860]$\\
            &$(0.05,0.01)$&$[-0.0189,+0.0134]$&$[-0.0176,+0.0161]$&$[-108.65,+77.034]$&$[-101.17,+92.556]$ \\
            \hline
            &$(0.0,0.0)$&$[-0.0025,+0.0021]$&$[-0.0068,+0.0054]$&$[-14.372,+12.072]$&$[-39.092,+31.043]$ \\
            $300$&$(0.02,0.005)$&$[-0.0130,+0.0082]$&$[-0.0123,+0.0108]$&$[-74.734,+47.140]$&$[-70.710,+62.087]$ \\
            &$(0.05,0.01)$&$[-0.0181,+0.0127]$&$[-0.0168,+0.0153]$&$[-104.05,+73.010]$&$[-96.580,+87.957]$ \\
            \hline
            &$(0.0,0.0)$&$[-0.0007,+0.0007]$&$[-0.0040,+0.0027]$&$[-4.0241,+4.0241]$&$[-22.995,+15.521]$ \\
            $3000$&$(0.02,0.005)$&$[-0.0127,+0.0070]$&$[-0.0120,+0.0105]$&$[-73.010,+40.241]$&$[-68.986,+60.362]$\\
            &$(0.05,0.01)$&$[-0.0180,+0.0126]$&$[-0.0167,+0.0152]$&$[-103.47,+72.435]$&$[-96.005,+87.382]$\\
            \hline
	\end{tabular*}
\end{table*}
 The table indicates a significant improvement in the limits of the anomalous couplings over the increment of luminosity in the ideal case of zero systematic errors. The limits improve by roughly a factor of 2 (1.4) when the luminosity increases from 35.9 fb$^{-1}$ to 150 fb$^{-1}$; They improve further by roughly a factor of 4 (2) when the luminosity is increased to 3000 fb$^{-1}$ for the anomalous coupling $\hat{\mu}_t$ ($\hat{d}_t$). However, when systematic errors are considered in the analysis, the limits on the couplings do not improve significantly when the luminosity is increased, as discussed earlier. In the specific case of systematic errors chosen as $(\epsilon_\sigma, \epsilon_A) = (0.02, 0.005)$, the limits on $\hat{\mu}_t$ and $\hat{d}_t$ only improve by a mere factor of approximately 1.2 going from the luminosity 35.9 fb$^{-1}$ to 3000 fb$^{-1}$.
This is because although the statistical error reduces when luminosity is increased, the systematic error becomes dominant, and the total error, comprising statistics and systematic errors, does not fall below the systematic error alone. When using a more conservative value for the systematic errors, such as $(0.05, 0.01)$, the improvement in the limits is even smaller over the increment of luminosity. The systematic errors have a significant impact on probing the $gt\bar{t}$ anomalous couplings in the $tW$ process and restrict the ability to obtain more precise measurements. 

A note on the impact of including higher-order electroweak (EW) effects on the limits of anomalous couplings is as follows. The electroweak corrections~\cite{Frixione:2008yi} are substantial, in addition to the QCD correction~\cite{Kidonakis:2021vob}, in the $tW^-$ production process. An interference effect is present between the $t\bar{t}$ process and the EW NLO $tW^-$ process accompanying an extra real $\bar{b}$-quark radiation. We estimated the EW correction in the SM without interference by removing the on-shell $t\bar{t}$ diagrams from the $tW^-\bar{b}$ process using the diagram removal technique~\cite{Frixione:2008yi} in {\tt Madgraph5}. The EW correction with interference is estimated by subtracting the $t\bar{t}$ cross~section from the $tW^-\bar{b}$ cross~section. To mimic our analysis with only one $b$-quark in the final state, we apply a veto of $p_T^{\rm veto}=20$ GeV to the extra $b$-jet. The above EW correction is estimated to be $19.2\%$ and $42.5\%$ with and without the above-mentioned interference, respectively, at the $13$ TeV LHC. The correction is sizable when compared to the $34\%$ QCD NLO corrections~\cite{Kidonakis:2021vob}. We obtain limits on the anomalous couplings, including the aforementioned EW NLO correction to the cross~sections, in addition to the QCD NLO effects, through the $k$-factor, and compare them against the limits with only QCD NLO corrections. We find that the limits on the anomalous couplings are improved with the inclusion of EW NLO corrections. For example, the limits on $\hat{\mu}_t$ and $\hat{d}_t$ are tightened by roughly $15\%$ with EW NLO corrections without interference, estimated at an integrated luminosity of $\mathcal{L}=3000$~fb$^{-1}$ and systematic uncertainties of ($\epsilon_\sigma,\epsilon_A) \in {(0.0,0.0),(0.05,0.01)}$. Conversely, the same limits improve by approximately $7\%$ when interference is taken into account.

 \section{Conclusion}
 \label{sec::con}
In this article, we studied the production of a single top quark associated with a $W$ boson in the fully leptonic final state. We investigated the anomalous chromo-magnetic and chromo-electric moments included in the $gt\bar{t}$ vertex, which receive contributions from physics beyond the SM. We utilized the polarization of the top quark and the $W$ boson and their spin correlations estimated from the angular distribution of the leptons, along with the production cross~section, to probe the anomalous couplings/moments. We reconstructed the two missing neutrinos using the MAOS method to obtain the rest frame of the top quark and the $W$ boson, enabling the measurement of polarization and spin correlation asymmetries. We estimated all the observables through a fully detector-level simulation and probe the anomalous couplings.

We studied the sensitivity of the cross~section, polarization asymmetries, and spin correlation asymmetries, binned over the reconstructed invariant mass of the top quark and the $W$ boson, to the anomalous couplings in both single-parameter and two-parameter variations. The polarization and correlation asymmetries significantly contribute to probing the anomalous couplings over the cross~section. The spin correlation asymmetries can also distinguish the $CP$ nature of anomalous couplings, unlike the polarization asymmetries. We estimated $95\%$ C.L. limits on the anomalous couplings for a set of integrated luminosity and systematic uncertainties. We observe that the limits do not improve substantially as the luminosity increases because the systematic uncertainty surpasses the statistical one for the $tW$ process.

Consequently, tightening the limits on the anomalous couplings at higher luminosity necessitates improving systematic uncertainty. The limits become loose considerably as the systematic errors increase. While the $tW$ process complements the $t\bar{t}$ process in the study of anomalous $gt\bar{t}$ couplings, our limits in the $tW$ process with a large $t\bar{t}$ background are somewhat tighter to the experimental limits obtained from the $t\bar{t}$ process~\cite{CMS:2019kzp} estimated for an integrated luminosity of $35.9$~fb$^{-1}$.

\begin{acknowledgments}
R.~Rahaman acknowledge the support from the Department of Atomic Energy, Government of India, for the Regional Centre for Accelerator-based Particle Physics (RECAPP), Harish Chandra Research Institute.	A.~Subba thanks the University Grant Commission, Govt. of India for financial support through the UGC-NET Fellowship.	
\end{acknowledgments}

\appendix
	
\section{Effect of binning the cross~section on $m_{tW}$ on the sensitivity to anomalous couplings}
    \label{Appendix1}
 \begin{figure*}[!t]
		\centering
		\subfigure{\includegraphics[width=0.49\textwidth]{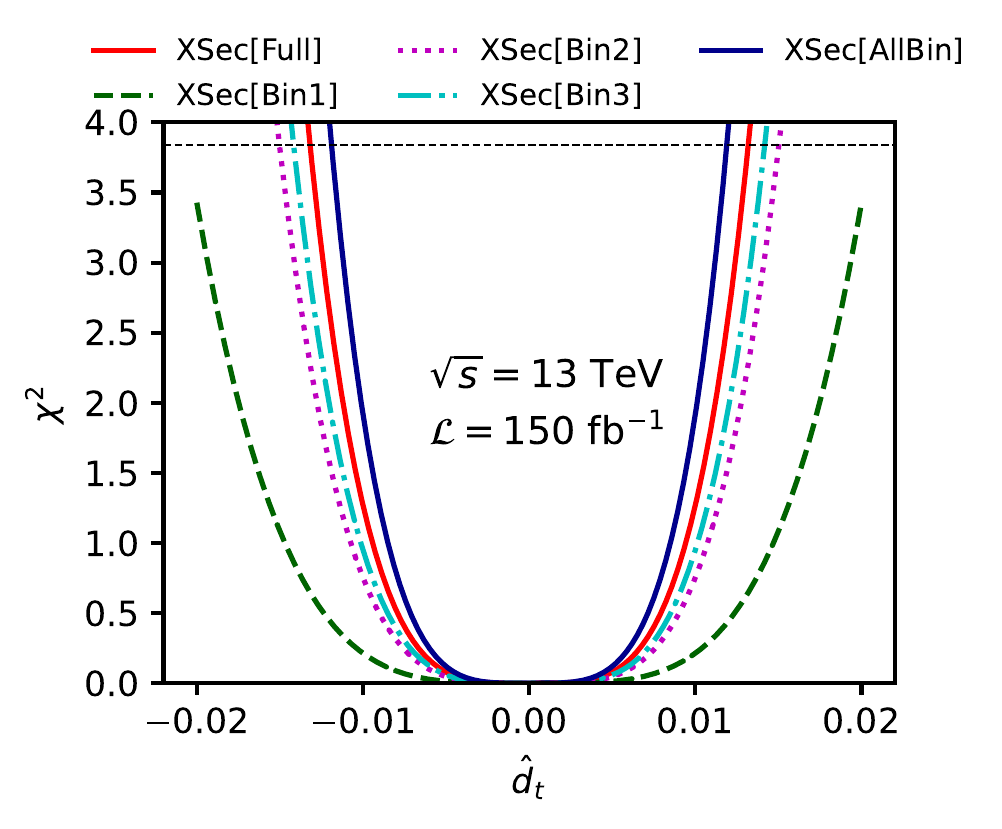}}
		\subfigure{\includegraphics[width=0.49\textwidth]{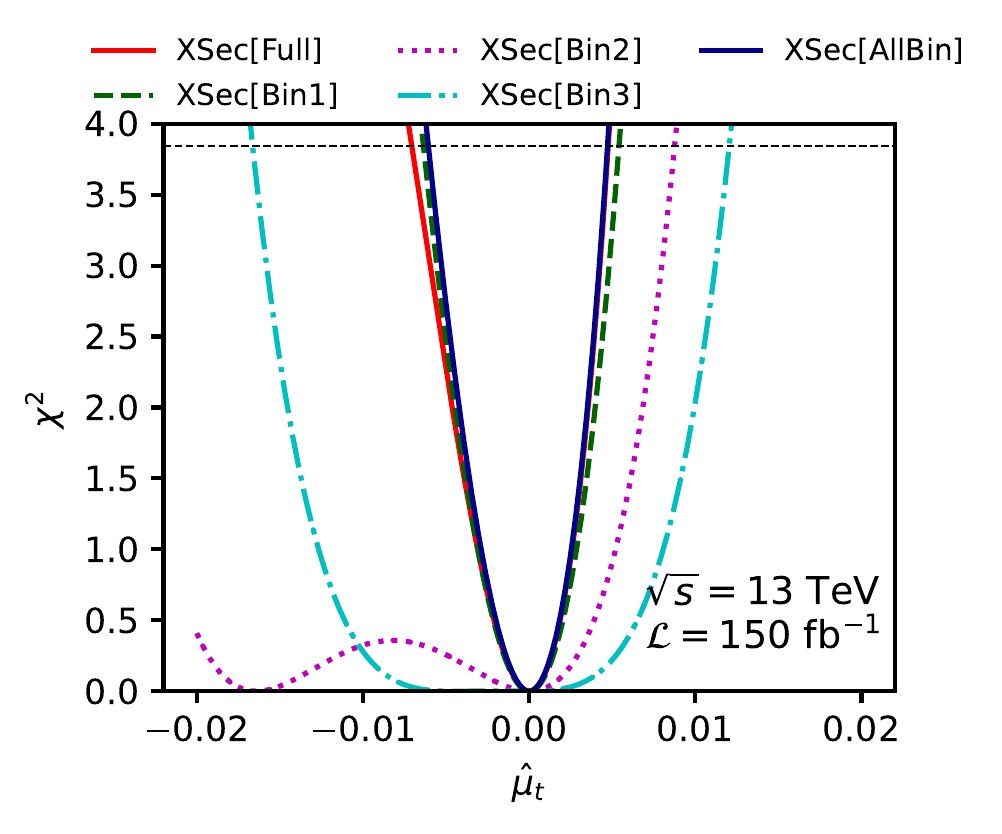}}
		\caption{\label{fig:xsec:binned} The $\chi^2$ function for cross~section as a function of anomalous couplings $\hat{d}_t$ and $\hat{\mu}_t$  in different bins of $m_{tW}$. The name of the bins are represented inside the third bracket in legend; Full and AllBin represent results without the binning and combination of all three bins, respectively.}
	\end{figure*}
We binned our observables into three intervals of $m_{tW}$ (see the distribution in Fig.~\ref{fig::kinematic}) to enhance the sensitivity to anomalous couplings. Here, we discuss the role of binning the cross~section in constraining the anomalous couplings estimated using the $\chi^2$ on binned and unbinned cases. The sensitivities in terms of $\chi^2$ are shown in Fig.~\ref{fig:xsec:binned}, using the cross~section in different bins, unbinned, and all combined bins. In the comparison, we choose $13$ TeV energy of the LHC and an integrated luminosity of $150$ fb$^{-1}$ with no systematic errors for simplicity. 
In the case of $\hat{\mu}_t$, which is a $CP$-even parameter, the cross~section in bin number one, denoted as Bin1 ({\em dashed/green} curve), provides a maximal contribution. The contribution of this bin alone is comparable to the contribution of the unbinned case ({\em solid/red} curve) denoted as Full in the legend. On the other hand, the other bins perform weakly in setting the limit for $\hat{\mu}_t$, and they provide an asymmetric limit.
In contrast, for the $CP$-odd moment $\hat{d}_t$, the contribution of the cross~section from bin numbers two and three is comparable to that of the unbinned case, while the cross~section from bin number one has the least contribution.
Finally, as expected, the limits set by the combination of all three bins are tighter than in the unbinned case. The binning also enhances the sensitivity of polarization and correlation asymmetries to the anomalous couplings, and they are used in the analysis.

 
\bibliography{refer}
	
\end{document}